\documentclass[twocolumn]{aastex63}

\newcommand{\solm}{M$_{\odot}$}

\usepackage{amsmath}
\usepackage{amssymb}
\usepackage[T1]{fontenc}
\usepackage{gensymb}
\usepackage{enumitem}
\usepackage{url}

\graphicspath{{./}{figures/}}

\received{January 1, 2018}
\revised{January 7, 2018}
\accepted{\today}

\submitjournal{ApJ}

\shorttitle{Observation of S4716}
\shortauthors{Pei$\beta$ker et al.}

\begin{document}

\title{Observation of S4716 - A star with a 4 year orbit around Sgr~A*}

\correspondingauthor{Florian Pei{\ss}ker}
\email{peissker@ph1.uni-koeln.de}
\author[0000-0002-9850-2708]{Florian Pei$\beta$ker}
\affil{I.Physikalisches Institut der Universit\"at zu K\"oln, Z\"ulpicher Str. 77, 50937 K\"oln, Germany}
\author{Andreas Eckart}
\affil{I.Physikalisches Institut der Universit\"at zu K\"oln, Z\"ulpicher Str. 77, 50937 K\"oln, Germany}
\affil{Max-Plank-Institut f\"ur Radioastronomie, Auf dem H\"ugel 69, 53121 Bonn, Germany}

\author[0000-0001-6450-1187]{Michal Zaja\v{c}ek}
\affil{Department of Theoretical Physics and Astrophysics, Faculty of Science, Masaryk University, Kotl\'a\v{r}sk\'a 2, 611 37 Brno, Czech Republic}



\author{Silke Britzen}
\affil{Max-Plank-Institut f\"ur Radioastronomie, Auf dem H\"ugel 69, 53121 Bonn, Germany}

\begin{abstract}

The ongoing monitoring of the Galactic center (GC) and Sgr~A*, the central supermassive black hole, produces surprising and unexpected findings. This goes hand in hand with the technical evolution of ground- and space-based telescopes and instruments, but also with the progression of image filter techniques such as the Lucy Richardson algorithm. As we continue to trace the members of the S-cluster close to Sgr~A* on their expected trajectory around the supermassive black hole, we present the finding of a new stellar source, which we call S4716. The newly found star orbits Sgr~A* in about 4.0 years and can be detected with NIRC2 (KECK), OSIRIS (KECK), SINFONI (VLT), NACO (VLT), and GRAVITY (VLTI). With a periapse distance of about 100 AU, S4716 shows an equivalent distance towards Sgr~A* as S4711. These fast-moving stars undergo a similar dynamical evolution, since S4711-S4716 share comparable orbital properties. We will furthermore draw a connection between the recent finding of a new faint star called S300 and the data presented here. Additionally, we observed a blend star event with S4716 and another new identified S-star S148 in 2017.
\end{abstract}

\keywords{editorials, notices --- 
miscellaneous --- catalogs --- surveys}

\section{Introduction} \label{sec:intro} 

Almost 5 decades ago, \cite{Lynden-Bell1971} proposed the existence of a supermassive black hole (SMBH) in the centre of our Galaxy. Following the claim by \cite{Lynden-Bell1971}, several generations of telescopes and instruments have revealed detailed information about the structure and components of the center of our galaxy. For example, observations with NACO \citep[][]{Lenzen2003, Rousset2003} that were mounted at the Very Large Telescope (VLT) have provided access to precise stellar measurements of single stars on their relativistic \citep[S2, see][]{Parsa2017} and non-relativistic \citep[S0-102, see][]{Meyer2012_s55} orbits around the central mass. This central mass can be described as the sum of the enclosed mass consisting of the compact mass (i.e. SMBH) and the extended mass (for example, nondetected faint stars, compact remnants, and dark matter).\\
From a historical point of view, \cite{Eckart1996_scluster} identified a few dozen stars orbiting Sgr~A* and named the related stellar members S-stars. These S-stars belong to the S-cluster and can be characterized by velocities up to several thousand km/s, with some of them moving on highly eccentric orbits \citep[][]{Ali2020}. It is a remarkable achievement on both the technological and scientific level to analyze stellar orbits that pass around Sgr~A* at distances comparable to the size of our solar system. Observations underline the multifaceted (star-formation) history of the Nuclear Star Cluster (NSC) and the embedded S-cluster \citep[][]{Lu2013, Schoedel2020, Nogueras-Lara2021}.\newline
The authors of \cite{Eckart1996_scluster} analyzed the S-stars using the Lucy-Richardson algorithm \citep[][]{Lucy1974} that stands for a high-pass filter. With these filtered data, \cite{Schoedel2002} was able to analyze for the first time the orbit of S2, the brightest member of the S-cluster with an orbital period of about 16 years. Furthermore, \cite{Ghez1998} and \cite{Eckart2002} suggested that the compact and variable radio source Sgr~A* \citep{Balick1974} can be associated with an SMBH. The unexpected finding of young O/B stars close to Sgr~A* resulted in the formulation of the "Paradox of Youth" \citep{Ghez2003} highlighting the uniqueness of the cluster. Due to the gravitational potential of Sgr~A* but also the high dust temperatures of about 200 K \citep{Cotera1999} any possible star formation process is hampered \citep{Jeans1902}.\newline
Due to a sensitive application of the high-pass filter in combination with a mild background subtraction, we are today able to observe stars that are even on shorter orbits than S2 \citep[mag$_K\approx$14, see][]{Rafelski2007} and S0-102/S55 \citep[mag$_K\approx$17, see][]{Meyer2012_s55}. These stars are S62 and S4711 with orbital periods of 9.9 years \citep[mag$_K\approx$16, see][]{Peissker2020a, peissker2021b} and 7.6 years \citep[mag$_K\approx$18, see][]{Peissker2020d}, respectively. A natural by-product of these short-orbit stars is the amount of the minimized enclosed mass, which results in higher precision regarding the calculation of the compact and extended mass.
Due to the confusion and blend stars \citep[][]{Sabha2012}, the observation of S62 and S4711 at a distance of a few mas from Sgr~A* can be highly challenging. On the other hand, stars at the edge of the S-cluster such as S29 are also not free of confusion. On their way towards the inner region of the S-cluster, the star interferes with several identified and non-identified S-stars causing a confused detection. 
From an observational point of view, the stellar confusion caused by close-by stars covers just one aspect. Due to the high extinction in the infrared of $A_{\lambda}\,=1-3$ \citep[][]{Schoedel2010, Fritz2011, Gautam2019} towards the Galactic center, a photometric analysis can be challenging, especially on large structures such as the NSC \citep[][]{schoedel2009}. However, substructures like the 40 mpc wide S-cluster should not be affected by extinction variations. The authors of \cite{Lu2013} for example use the common NIR K-band value $A_{K}\,=\,2.7$ for their analysis of the entire S-cluster.\newline
However, the motivation for the observation of S-stars is surely related to the supermassive black hole Sgr~A* and how stellar orbits evolve in the strong-gravity regime \citep{Parsa2017,2017PhRvL.118u1101H,2020PhRvL.124h1101H}. With a precise knowledge of the orbital parameters, one is able to determine properties of Sgr~A* which are necessary for theoretical models \citep[see, e.g.,][]{Zajacek2020, Tursunov2020, Sukova2021}. For example, the relativistic prograde pericenter advance of a star per its orbital period around a central mass can be determined by
\begin{equation}\label{adv}
\Delta\varphi=\frac{6\pi G}{c^2}\, \frac{M}{a(1-e^2)},
\end{equation}
where $M$ is the mass of Sgr~A* \citep[see][]{Weinberg1972}. The gravitational length scale is defined by
\begin{equation}\label{scale}
    r_{\rm g}=\frac{GM}{c^2}\,=\,1.96\,\times\,10^{-4}\,{\rm mpc}\sim 4.9 \mu{\rm as}\,,
\end{equation}
with $M\,=\,4.1\,\times\,10^6 M_{\odot}$, G the gravitational constant, and $c$ the speed of light \citep[see also][]{Rubilar2001}.
The semi-major axis $a$ and the eccentricity $e$ are inferred from the orbital solution of a given star. Furthermore, the relativistic parameter is given by $\Gamma\,=\,\frac{r_{\rm S}}{r_{\rm p}}$ where $r_{\rm S}$ represents the Schwarzschild radius of Sgr~A* with $r_{\rm S}=2GM/c^2=1.21\,\times\,10^{12}(M/4.1\,\times\,10^6 M_{\odot})$ cm and $r_{\rm p}$ represents the pericenter distance of the stellar probe with
\begin{equation}
    r_p\,=\,a(1-e)\,\,\, .
\end{equation}
Comparison of relativistic parameters in units of $10^{-4}$ for S2 (6.8), S62 (46), S4711 (5.6), and S4714 (64) \citep{Peissker2020d} in relation to their related K-band magnitude reveals the challenging framework of the observations. Although S62 and S4714 are surely more suitable to investigate the relativistic influence of Sgr~A*, the faint emission in combination with the influence of the dominating point spread function (PSF) of S2 hinders a confusion-free detection. Moreover, common blend star scenarios as well as crowding problems caused by nearby/background stars are increasing the uncertainty range. One has furthermore take into account a stellar density of over $\rm N={10.0}/arcsec^2$ in a distance of 40 mpc (i.e., the size of the S-cluster) for sources with a K-band magnitude of $\rm m_K\,=\,17.5-18.5$ \citep[][]{Gallego-Cano2018}.\newline
By inspecting the results of \cite{Gillessen2009}, \cite{Meyer2012_s55}, \cite{Parsa2017}, \cite{Gillessen2017}, and \cite{Peissker2020a, Peissker2020d} we come to the conclusion that using the iterative Lucy-Richardson process is the most reliable way of analyzing data of a crowded area, such as the S-cluster. For example, the well-investigated orbit of S55/S0-102 can be revealed using a high-pass filter, such as the Lucy-Richardson algorithm, because overlapping PSF wings suppress star emission.\newline
This approach can be applied to most of the S-cluster members. Most recently, the orbit of the star S29 gained attention because of a disputed detection between 2015 and 2021 \citep[][]{Peissker2020a, GravityCollaboration2021, peissker2021b}. This confusion resulted in the observation of the periapse of a star in 2021 that was interpreted as S29 by \cite{GravityCollaboration2022a}. Since we already showed the trajectory of S29 in \cite{peissker2021b} on its trajectory towards Sgr~A*, we will 
present an alternative interpretation with the introduction of a new S-star S4716. We chose this name to follow up on our nomenclature presented in \cite{Peissker2020d} where we introduced several new stars, which we call S4711-S4715. These stars are orbiting Sgr~A* on highly eccentric orbits with periods close to or less than 10 years. For S4716, we derive an orbital period of only 4 years. With a distance of about 12 mas during its periapse, S4716 has the shortest known orbital period around a supermassive black hole to date.

\begin{figure}[htbp!]
	\centering
	\includegraphics[width=.5\textwidth]{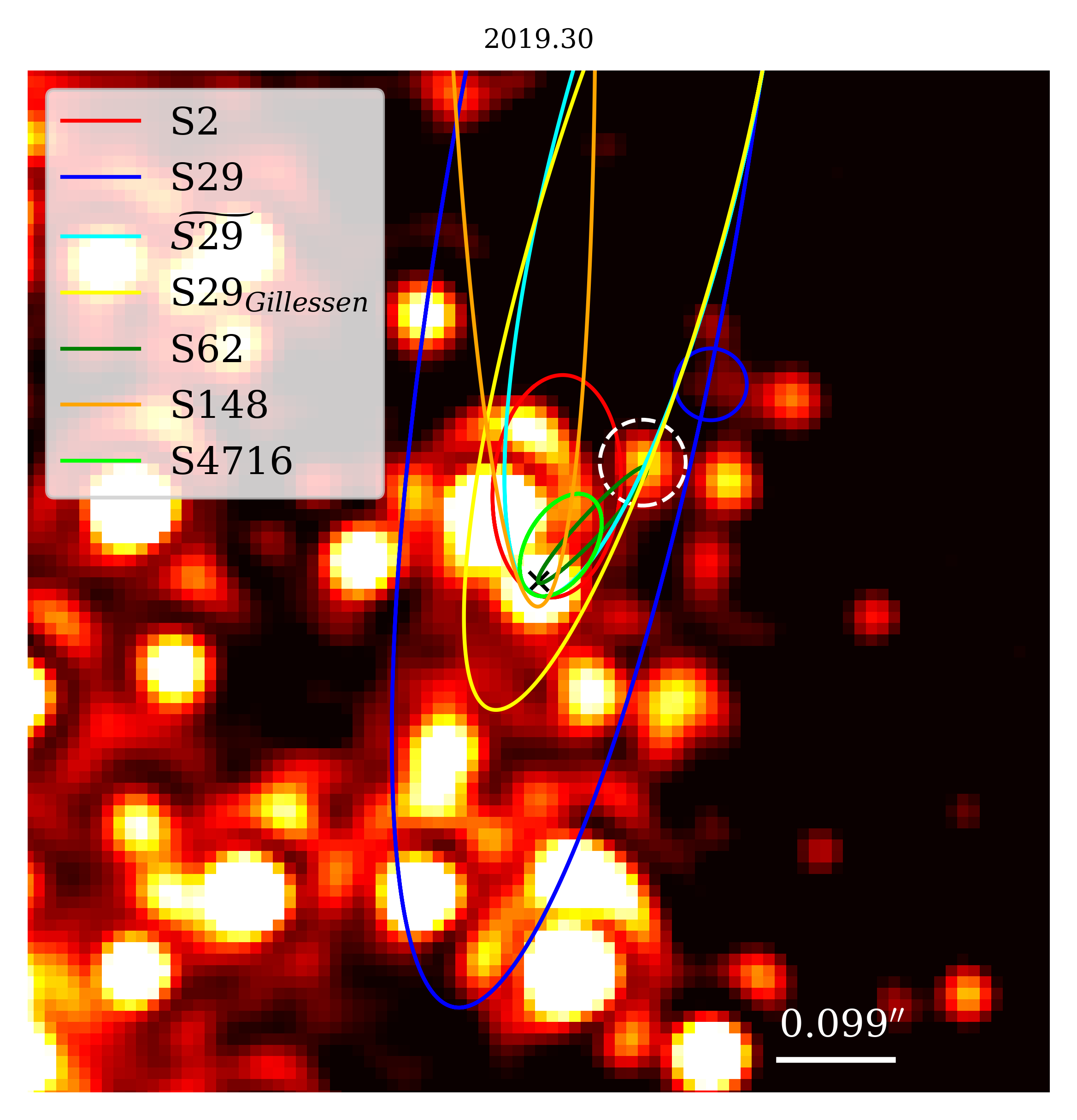}
	\caption{The K-band view of the Galactic center observed with NIRC2 (KECK) in 2019.30. This image is high-pass filtered and shows the position of several S-stars close to Sgr~A* which indicated by a black $\times$. The star in the white dashed circle shows a K-band magnitude of $16.3$ mag while the star in the blue circle is $17.0$ mag faint. Both K-band magnitudes are consistent with the reported values for S62 by \cite{Peissker2020a} and S29 by \cite{Gillessen2017} and \cite{peissker2021b}, respectively. Here the north is up and the east is to the left.}
\label{fig:finding_chart_2019}
\end{figure}

The paper is structured as follows.
In Sec. \ref{sec:data}, we will provide a brief overview of the data used (see also Appendix \ref{sec:data_appendix}) and the methods. We furthermore compare different analyzing techniques. Then we present the results of our analysis in Sec. \ref{sec:results} and discuss the results in Sec. \ref{sec:discussion}. Our final conclusions of the here presented data can be found in Sec. \ref{sec:conclusion}. In the Appendix, we list the used data, provide further results, and propose a guideline for the detection of single stars in a crowded stellar field. In addition , we will introduce another new S-cluster member called S148. We will observe the emergence of a blend star event that justifies the checklist at the end of this work.
In the following, we will focus on six stars: S148 (this work), S4716 (this work), S29 \citep[see][]{peissker2021b}, S62 \citep[][]{Peissker2020a}, S29$_{\rm Gillessen}$ \citep[see][]{Gillessen2017}, and $\rm \widetilde{S29}$ \citep[see][]{GravityCollaboration2022a}. See the finding chart in Fig. \ref{fig:finding_chart_2019} where we indicate the position of the latter three in the white dashed circle.
We will abbreviate \cite{Gillessen2017} by G17, \cite{GravityCollaboration2021} with GColl2021a, and \cite{GravityCollaboration2022a} with GColl2022a. Since we are aware of possible confusions for the reader, all abbreviations used in this work are listed in Table \ref{tab:abbreviations}.
\begin{table*}[htb]
\centering
\begin{tabular}{|c|c|c|}\hline \hline
Star  &  related Publication &	abbreviated Publication  \\ \hline 
 S148   & this work & - \\ \hline
 S4716  & this work & - \\ \hline
 S29    & \cite{peissker2021b} & - \\ \hline 
 S62    & \cite{Peissker2020a} & - \\ \hline
 S29$_{\rm Gillessen}$  & \cite{Gillessen2017}  & G17 \\ \hline
 $\rm \widetilde{S29}$  & \cite{GravityCollaboration2022a} & GColl2022a \\ \hline
    & \cite{GravityCollaboration2021} & GColl2021a \\ 
\hline \hline
\end{tabular}
\caption{Abbreviations used in this work. We list the investigated stars in this work and the related publications that should be consulted by the interested reader.}
\label{tab:abbreviations}
\end{table*}

\section{Data and Analysis} 
\label{sec:data}
Here we provide an overview of the data reduction and a short introduction of the used instruments. In addition, we list the tools that we have used to obtain our final images. We will furthermore describe the applied background subtraction because of the dominating background in the S-cluster, especially close to Sgr~A*. The used data is listed in the Appendix \ref{sec:data_appendix}.

\subsection{SINFONI}

The Spectrograph for INtegral Field Observations in the Near Infrared (SINFONI) was mounted up until 2019 at the Very Large Telescope (VLT) at the European Southern Observatory (ESO) \citep[][]{Eisenhauer2003, Bonnet2004}. It is now decommissioned. The instrument contained an Integrated Field Unit (IFU) which is responsible for the three-dimensional shape of the data (two spatial dimensions in $x$ and $y$ directions, and one spectral dimension along the $z$ axis). With this setup, every pixel has a related spectrum. Continuum images and line maps covering the H+K band can be extracted from the data cube. In this work, we limit the positional analysis to K-band continuum images ($2.0\,-\,2.2\,\mu m$).\newline
For the data reduction, we used the ESO pipeline with DARK-, FLAT-, LINEARITY-, WAVE-, and DISTORTION-correction files. The final single 3d data cubes are then cropped to exclude non-linear edges and shifted in an array to a specific position $(x_i,\,y_i)$. This position is determined by a manually created finding chart. Single data cubes ($64\,\times\,64$ pixel) have a spatial pixel scale of 12.5 mas resulting in a field-of-view (FOV) of $0.8\,\times\,0.8$ arcsec with an integration time of 600 seconds. The final data cubes have a FOV of about $1.2\,\times\,1.2$ arcsec. The stacked integration time depends on the number of single data cubes. Adaptive optics (AO) for every observation is enabled and, if possible, used with an optical natural guide star in the proximity of Sgr~A*.  

\subsection{NACO}

The Nasmyth Adaptive Optics System (NAOS) Near-Infrared Imager and Spectrograph (CONICA), abbreviated to NACO, was decommissioned in 2019 at the VLT \citep[][]{Lenzen2003, Rousset2003}. The imager operated in the H-, K-, L-, and M-band with an infrared wavefront sensor for AO. For the nonarchival observations, we used a $5.0\,\times\,5.0$ arcsec box with a random dither pattern. For every here analyzed observation, the bright pulsating supergiant IRS7 located $\sim 5.5"$ north of Sgr~A* is used for the AO correction. The final images are shifted and added to increase the on-source integration time. The spatial pixel scale of the here analyzed data is 13.3 mas with a FOV of $1024\,\times\,1024$ pixels. The observations were executed in the $\rm K_s$-band that is covering $1.97-2.32\mu m$.

\subsection{NIRC2}

At the KECKII telescope (located at Hawaii/USA), the Near-Infrared Cam2 (NIRC2) is mounted \citep[][]{Matthews1994, Nelson1997}. The imager operates in the H-,K-, and L-bands using AO to correct for atmospheric turbulence. The FOV of single images is $1024\,\times\,1024$ pixel with a spatial pixel scale of 9.9 mas. From the KECK Observatory Archive (KOA\footnote{\url{https://koa.ipac.caltech.edu/}}), we downloaded the public data that was observed in the $\rm K_p$-band covering the wavelength range $1.94-2.29\mu m$.

\subsection{OSIRIS}

Like SINFONI, the OH-Suppressing Infrared Imaging Spectrograph (OSIRIS) produces 3d data cubes with 2 spatial and 1 spectral dimension \citep[][]{Larkin2006, Mieda2014}. The KECKII mounted instrument uses an Integrated Field Spectrograph (IFS) with AO. The instrument itself also contains a science camera that produces continuum images of the related scientific object. In this work, we will focus on the results of this science camera. The OSIRIS imager has a FOV of $2048\,\times\,2048$ pixel. Here, the presented data is observed with the Kn3 filter covering the K-band from about $2.12\,\mu m$ to $2.22\,\mu m$ with a spatial pixel scale of 10 mas. We shifted and added the downloaded data from the KOA to increase the on-source integration time to about 8 seconds. 


\subsection{High-pass filter and methods}

In general, high-pass filters are used to suppress background noise and dampen the wings/airy rings of a PSF. Since objects can move through intersecting regions that are polluted by overlapping airy rings or wings, one can reduce this influence by a mild background subtraction. This process is not only needed for the analysis of a crowded region, but it maximizes the validation of the data. Since artefacts are removed by the background subtraction, only emission above the noise level is observed. Vice versa, if the image is dominated by noise, every stellar emission is suppressed and hence, no object can be observed. 
\begin{figure*}[ht!]
	\centering
	\includegraphics[width=1.\textwidth]{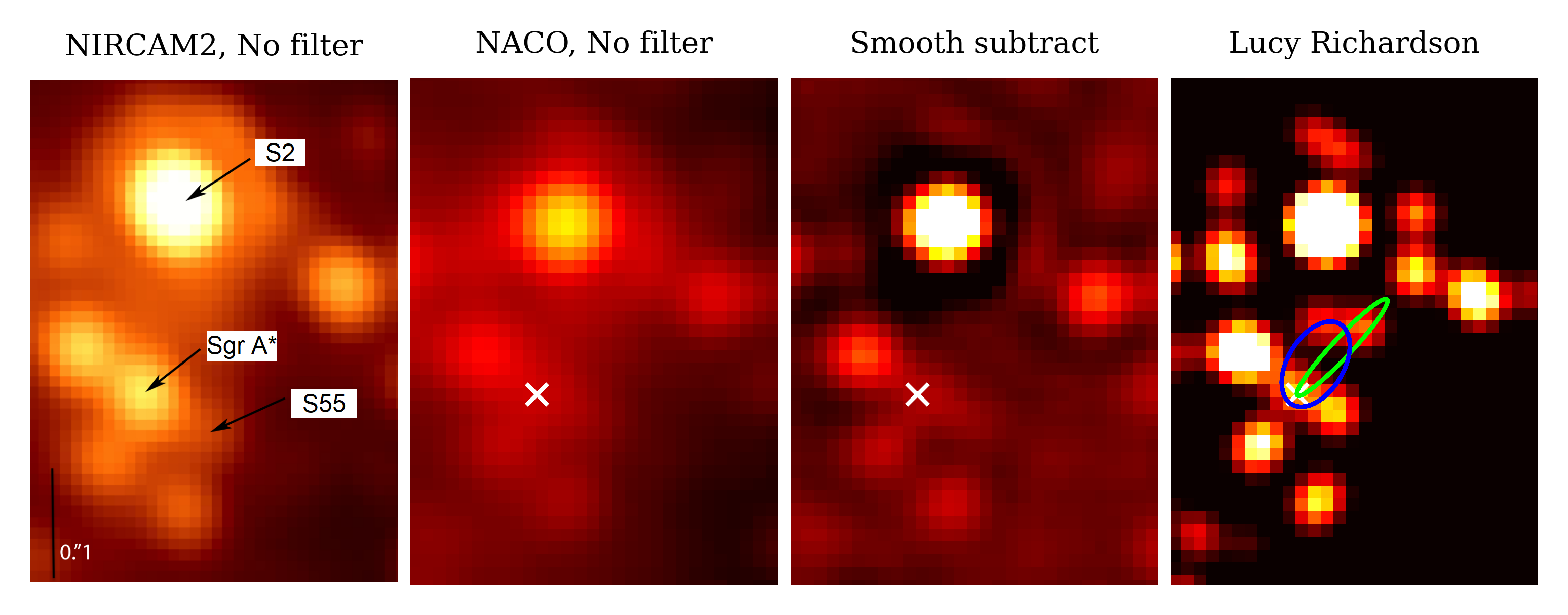}
	\caption{Comparison of the K-band continuum emission with two different instruments and two high-pass filter techniques in 2010. The left NIRCAM2 image is adapted from \cite{Meyer2012_s55} while all the other plots show NACO data. In the far right image, we indicate the orbit of S62 and in this work discussed new S-cluster member S4716 with lime- and blue-colored trajectories, respectively. Please compare the observation of S55 in every presented panel. In every image, north is up, east is to the left.}
\label{fig:filter}
\end{figure*}
In this work, we use two different techniques, namely the smooth subtraction (SM) and the Lucy-Richardson (LR) algorithm, with the focus on the latter. While the SM-algorithm is a reliable tool, the resulting images still suffer from noise and extended diffraction minima that can be observed especially around S2. On the other hand, the LR-algorithm is sensitive to the chosen input parameters and should be complementary to the SM-algorithm to avoid a confused analysis. Even though we described the two filter techniques numerous times \citep[see, e.g., ][]{Peissker2020a, Peissker2020d, peissker2021b}, we will list the necessary steps for transparency in the following. In Fig. \ref{fig:filter}, we display K-band images observed with NIRC2 (KECK) and NACO (VLT). We compare these continuum images with two different high-pass filtering techniques, namely, the SM- and LR-alogrithms. Figure \ref{fig:filter} also shows the difficulties of observing and finding new and unknown stellar S-cluster members. Hence, we will provide a checklist in Appendix \ref{appendix:checklist} that establishes a frame for the identification of unknown sources.

\subsubsection{Smooth-subtract algorithm}

As an advantage, the SM-filter is easy to use and handle. The final image $\rm I_{all}$ is a convolution of point sources and a PSF. The final image $\rm I_{all}$, however, is a combination of all observed frequencies whereas a smoothed version of $\rm I_{all}$ is representing a low-pass filtered image $\rm I_{low}$. To derive the final SM-filtered image, we can write
\begin{equation}
    \rm I_{all}\,-\,\rm I_{low}\,=\,\rm I_{high}
\end{equation}
where $\rm I_{high}$ is the desired result. With the above equation, it becomes eminent that the only input parameter of this procedure is the smoothing PSF that we denote with $\rm PSF_{smooth}$. As a general procedure, one set $\rm PSF_{smooth}\,=\,PSF_{Gauss}$ where $\rm PSF_{Gauss}$ represents a Gaussian filter with a size that matches the PSF of the data. As it is mentioned before, the subtraction of a PSF often results in a negative flux ring around the brightest source, making it impossible to analyze the close-by regions. 

\subsubsection{Lucy-Richardson algorithm}
\label{sec:filter}
As a major keystone, the PSF and background subtraction for the LR-filter determines the output and should be used as a complementary process or be cross-checked with different datasets. Nevertheless, there are different procedures available to extract a PSF \citep[see for example, StarFinder,][]{Diolaiti2000}. We will focus here on a manual approach to have full control over the parameters. As a first step, we check isolated stars in the crowded region. Here, a star is labeled {\it isolated} if the distance to the next near-by source is at least one PSF. Possible reference stars are, for example, S4, S6, S7, S30, S65 at a distance of 0.2-0.8 arcsec from Sgr~A* \citep[][]{Ali2020}. With the selection of these local stars, we minimize anisoplanatic effects \citep[][]{schoedel2010b}. 
Since this approach cannot always be fulfilled (for example, the SINFONI FOV is highly limited in size), at least S2 can be used to get the first idea of the true PSF. The parameters are derived by fitting different-sized Gaussians to the stars. Common parameters are the $x$ (FWHM1) and $y$ (FWHM2) values as well as the angle of the PSF. Another obvious parameter is the magnitude of the reference stars. With $\rm mag_K\,=\,14.1$ mag, S2 is the brightest S-cluster member and thus is the most prominent star in the data investigated. Hence, the reference stars should not be too faint, i.e., too close to the background since this could affect the shape of the PSF. A K-band magnitude of $\leq\,15.5$ mag is a reliable choice since the background emission accounts for about 2-5\% of the peak flux of the reference source.
The next step is the application of the LR-algorithm to get rid of the dominant side lobes of the image PSF. After this deconvolution step, the resulting delta function map is convolved with a PSF with the size dimensions of $\rm =\,(FWHM1+FWHM2)/2$. In the resulting convolved image, the brightest and isolated sources are again fitted with a Gaussian of different sizes if the image PSF is not circularly shaped. 
\begin{figure}[ht!]
	\centering
	\includegraphics[width=.5\textwidth]{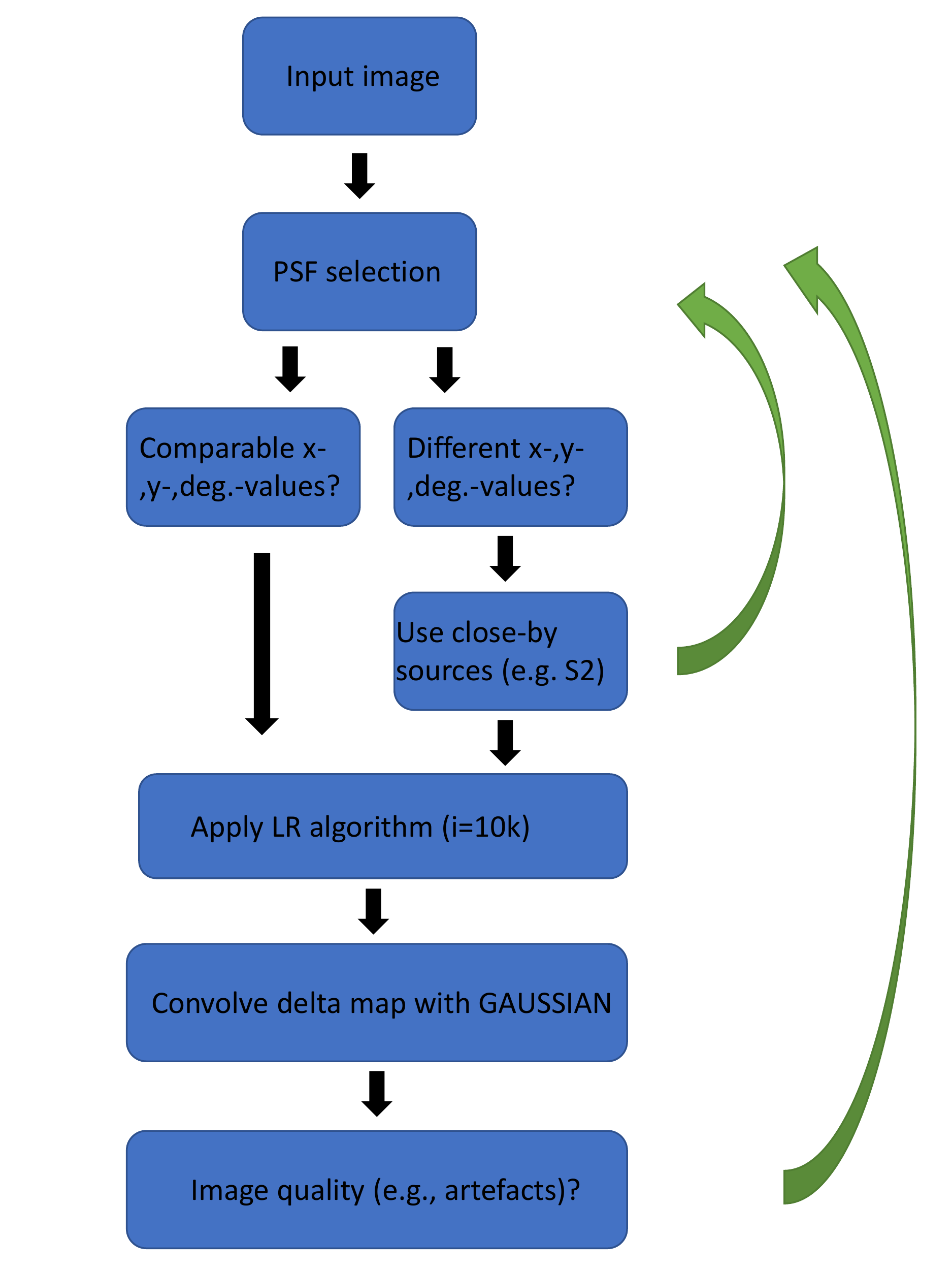}
	\caption{Flowchart of iterative process for using the LR algorithm. The input image is a FITS file. However, other processable formats are allowed. The PSF selection is usually done outside the S-cluster because of the number of isolated sources (see text for details). A FOV can suffer from a variable PSF which increases the challenge of finding a suitable solution. However, we limit the analysis to a small FOV where we assume, that the PSF is stable to first order (i.e., variations of less than 0.25 pixel are allowed). Because S2 is usually not an isolated source, it should only be used if no other stars can be found (e.g., the SINFONI data). After applying the LR algorithm with iterations of 10k, we convolve the resulting delta map with a GAUSSIAN. The convolved image is the solution if the image does not suffer from artefacts. The process can be cross-checked by the position of, for example, S2.}
\label{fig:isketch}
\end{figure}
This process will be repeated until the fitted PSF converges to a circular shape with a minimized elongation. The methodology of this process is shown in Fig. \ref{fig:isketch}. We note that the process described with the flow diagram shown in \ref{fig:isketch} only serves as an overview. If, for example, the overall quality of the data are reduced, this iterative will not result in the finding of a suitable PSF.\newline
Although background subtraction can help to find the real PSF of the image, a much more applicable approach is the construction of an Airy-ring filter that we call $\rm PSF_a$. Once the iterative process described here is complete, a suitable PSF is constructed which we denote by $\rm PSF_{psf}$. We create a new filter with
\begin{equation}
    \rm PSF_{a}\,\otimes\,PSF_{psf}\,=\,PSF_{apsf}
\end{equation}
where $\rm PSF_{aspf}$ denotes the convolved result of the input. Using the LR-algorithm with $\rm PSF_{aspf}$ in combination with a background subtraction, we derive robust results showing stars at their expected orbital positions. We note that the process described is the same for every data set.

\subsubsection{Sophisticated background subtraction}

By analyzing the S-cluster, one is permanently confronted with the dominant and variable background \citep[][]{Tep2021}. One example that underlines the complexity of the topic is the faint star S300 that is observed with GRAVITY (GColl2022a). The existence of S300 with its apparent K-band magnitude of almost 20 mag adds to the cusp of faint stars that are under the detection limit regarding the imaging data observed with, for example, SINFONI, NACO, or NIRC2.\newline
As shown by \citet{Sabha2012}, the diffuse background emission increases towards Sgr~A*. Because of the structure of the S-cluster, the increasing/decreasing background is not symmetrical with respect to Sgr~A*. For the moment, we will ignore the nonuniform shape of the cluster.
Assuming an image without errors (e.g., dead/hot pixels or cosmic rays), we can define every pixel as
\begin{equation}
    \rm signal_{obs}\,=\,signal_{obj}\,+\,signal_{background}
\end{equation}
where $\rm signal_{obj}$ is related to an arbitrary object (e.g., a star). Naturally, $\rm signal_{obj}$ can be equal to zero and the related pixel information is limited to the background emission. Since the variable background is a function of the position $r$, it is 
\begin{equation}
    \rm f_{(r)}\,=\,signal_{background}
\end{equation}
and 
\begin{equation}
    \rm -grad(f_{(r)})\,\neq\,0
\end{equation}
where $\rm grad(f_{(r)})$ is directed towards Sgr~A*. We define $z\,=\,f_{(r)}\,=\,x\cdot y$ and set $\rm R.A.\,=\,x$, $\rm DEC\,=\,y$. We use 
\begin{equation}
    \rm -grad(f_{(r)})\,=\,(\frac{\partial f}{\partial r_i}(r),....,\frac{\partial f}{\partial r_n}(r))\,=\,\sum_i \frac{\partial f}{\partial r_i}(r)\vec{e_i}
\end{equation}
and since we observe the on-sky projected view, we limit the above equation to $i\,=\,z$. The modulus yields $y\,=\,\frac{1}{x}$, ignoring negative values. Using now $\rm signal_{obj}\,\neq\,0$, we can investigate different scenarios for the background subtraction procedure.
\begin{figure}[ht!]
	\centering
	\includegraphics[width=.5\textwidth]{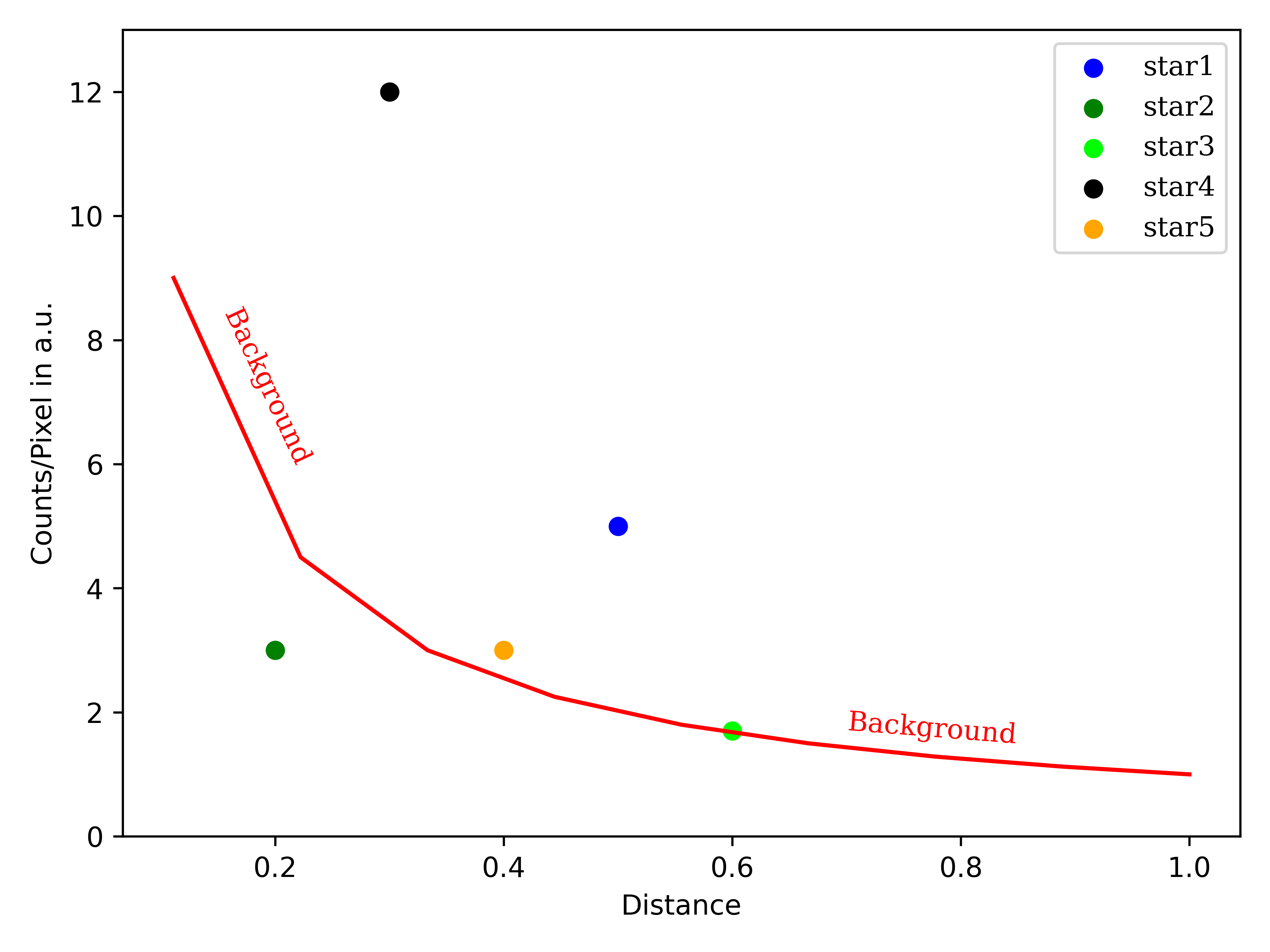}
	\caption{A trend of the background inside the S-cluster. Here we normalize the size of the S-cluster to unity. Please note that this trend can differ depending on the direction because of the non-uniform shape of the S-cluster. Observations of this trend are shown in \cite{Sabha2012} and \cite{peissker2021c}.}
\label{fig:bs}
\end{figure}
In Fig. \ref{fig:bs}, we show the trend for the diffuse background in an arbitrary direction where the origin is defined as the position of Sgr~A*. As we pointed out before, the shape of the S-cluster is nonuniform and hence, the trend in Fig. \ref{fig:bs} might differ for different directions. As a consequence, we introduce a variable parameter ``b'' with $\rm b\,\in\,\mathbb{R^+}/\{0\}$ and $y\,=\,\frac{1}{x}\cdot b$. We set $b=1$ for the example in Fig. \ref{fig:bs} and randomly disperse 5 stars with different magnitudes. These represent different background subtraction cases:
\begin{itemize}
    \item{star 1:} The star is above the background and can be observed. If, however, star 1 moves closer towards Sgr~A*, the stellar magnitude might increase. For a confusion-free detection, a background subtraction can be applied. One should note that an adaptive background subtraction could result in a decreased chance to observe star 5. The stellar object star 4 can be observed regardless of the background subtraction.  
    \item{star 2:} This star cannot be observed at this distance but contributes to the variable background emission. It can be assumed that this star also orbits Sgr~A* and creates fluctuations in the background or a blend-star scenario. It is important to note that this star can be observed if the defined orbit allows a sufficient distance with respect to Sgr~A*. 
    \item{star 3:} The star is confused with the background and cannot be observed at this distance. It can be treated like star 2.
    \item{star 4:} With a magnitude above the highest noise level, this star can be observed regardless of the background or the distance to Sgr~A*. Of course, if another star with a comparable magnitude is close by, the confusion is increased.
    \item{star 5:} Observing star 5 is possible but challenging because of the prominent noise (e.g., S55 and S62). Again, an increased distance with respect to Sgr~A* enhances the probability to observe a star like star 5.
\end{itemize}
We refer the interested reader to the Appendix of \cite{Peissker2020a} where we consider additional aspects of the background emission and source crowding.

\subsection{Methodology}
\label{sec:methodology}
Here, we outline the process of extracting data points from our data set. Since we use data from 5 different instruments that are mounted at VLT and KECK, we have to take into account possible small offsets between the different reference frames. The general problem is described \cite{Gillessen2009S2} where the authors show small offsets between the derived orbital solution for S2/S0-2 using KECK and VLT data. In contrast, we cannot confirm the given offsets in \cite{Gillessen2009S2} of $\Delta x\,=\,-3.7\pm 0.6$ mas and $\Delta y\,=\,-4.1\pm 0.6$ mas. Using a 17 year data baseline to derive orbital solutions for S-stars like, for example, S29 observed with NACO and SINFONI, we find no significant offset. As we show in this work, the orbital solutions are interchangeable. As discussed in \cite{peissker2021c}, the background noise is the most dominant contributing factor to uncertainties.\newline
Matching Keplerian stellar solutions covers one aspect of the data treatment. In addition to noise (caused by the background or the instrument), the exact position of Sgr~A* is the another strong source of uncertainties. As described in \cite{Gillessen2009} and \cite{Parsa2017}, the SiO maser stars in combination with a linear transformation is a suitable way to derive the position of Sgr~A*. This is reflected in a precise determination of the orbital elements of S2 allowing to investigate, for example, the Schwarzschild precision. In agreement with \cite{Gravity2018}, the authors of \cite{Parsa2017} demonstrated that the NACO data provide a high level of precision. Furthermore, the general process for deriving the position of Sgr~A* in the GRAVITY and NACO data is comparable and results in the matching outcomes as the authors mentioned above find a relativistic parameter of 0.88 \citep[][]{Parsa2017} and 0.9 \citep[][]{Gravity2018}. The VLT results are accompanied by the analysis of \cite{Do2019_S2}, where the authors estimate a relativistic parameter of 0.88 for S0-2/S2 observed with the KECK telescope. Therefore, we can conclude that the well-observed orbit of S2 already includes several detector and reference frame corrections. Hence, we use the orbital elements of S2 from \cite{Do2019_S2} and GColl2022a for the analysis of the KECK and VLT data, respectively.

\section{Results} \label{sec:results}
Here we present the results of our analysis. We introduce S4716, a new member of the fast and faint S-cluster population \citep[FSS-population, see also][]{Peissker2020d}.

\subsection{The Keplerian orbit of S4716}

As we constantly analyze the existing and new imaging data of the GC, we inspected the data presented in \cite{Peissker2020a}, \cite{Peissker2020d}, and \cite{peissker2021b} with the here presented fine-tuned approach. Based on this analysis, we recognized an unidentified source between Sgr~A* and S2\footnote{Please consult \cite{Peissker2020d} and more recently, \cite{peissker2021b}.}. Especially the data of 2007 and 2019 showed a bright source that does not fit in our stellar orbit list \citep[][]{Ali2020}. 
\begin{figure*}[htbp!]
	\centering
	\includegraphics[width=1.\textwidth]{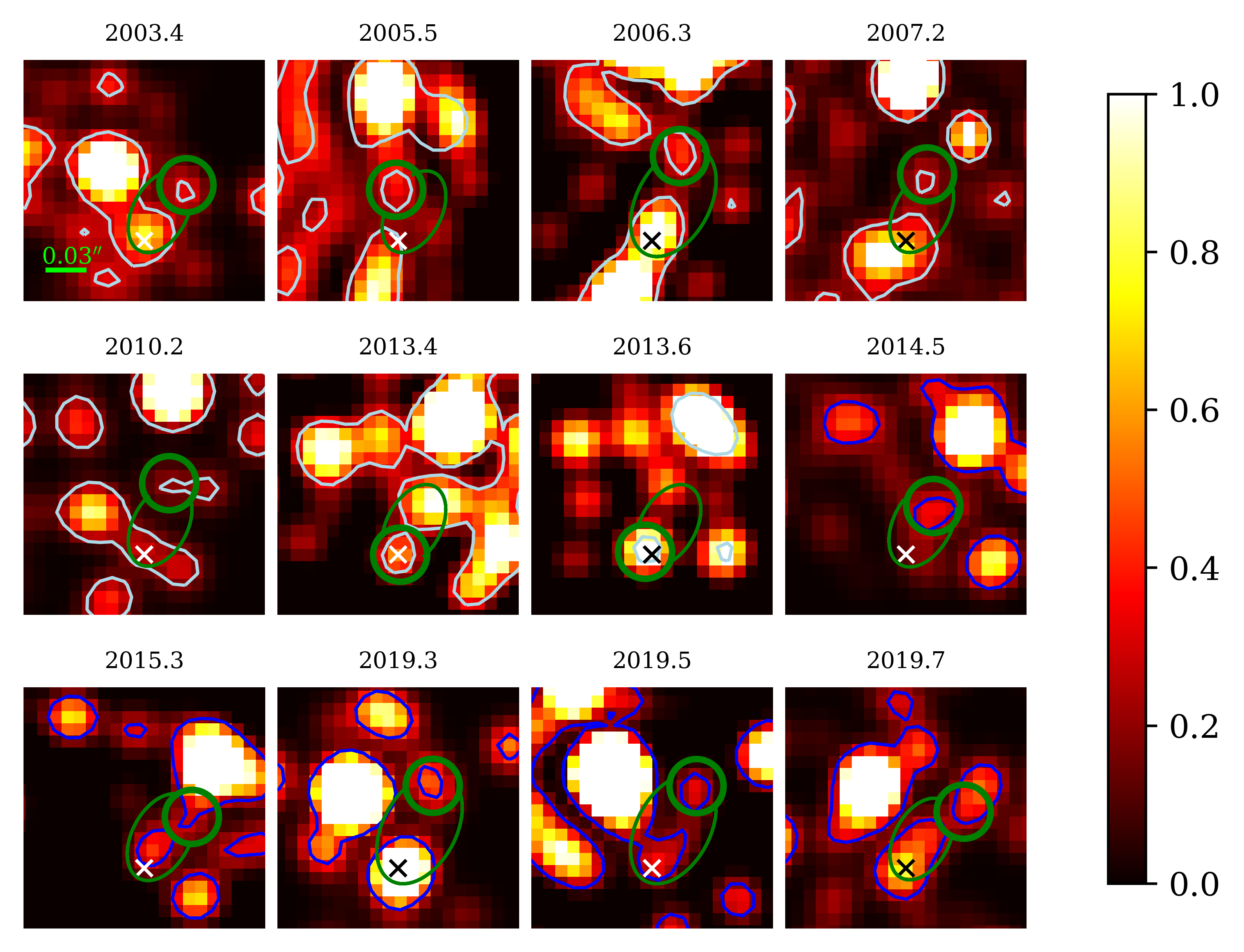}
	\caption{Timeline of the new identified star S4716 on its Keplerian orbit around Sgr~A*. Here we show 12 epochs between 2003 and 2019 in the direct vicinity of Sgr~A*. The data of 2006.3, 2019.3, and 2019.5 are observed with NIRC2/KECK. In 2013.4, 2014.5, and 2015.3, we show observations executed with SINFONI/VLT. All other epochs are observed with NACO/VLT. Sgr~A* is indicated by a cross with S4716 at the expected position marked with a green circle. We want to underline that for every shown epoch, S62 can be found at the expected position based on the orbital elements presented in \cite{Peissker2020a} (except for 2015 where S62 is confused with S2). Therefore, we present exactly the same figure in the Appendix with the orbit of S62. Here, north is up and east is to the left. Every panel shows a 20$\times$20 pixel image of the direct vicinity of Sgr~A* and is normalized to the peak emission of S2 (which corresponds to a K-band magnitude of 14.1). The contrast is choosen to show also fainter sources that may seem brighter than the estimated magnitude.}
\label{fig:timeline}
\end{figure*}
Between 2003 and 2020, we identify this stellar candidate in 16 epochs (see Table \ref{tab:positions_instruments_band}, Appendix \ref{sec:data_appendix} and Fig. \ref{fig:timeline}) for a selected overview). Limitations are the stellar interference from various sources such as, for example, S2, S55, and S62. This can be observed, for example, in 2004, 2011, and 2012 where S62 \citep[][]{peissker2021b} complicates a confusion-free observation\footnote{See the subplot for 2010.2 in Fig. \ref{fig:timeline}. The star on the right next to S4716 is S62.}.
In addition, the configuration of the Fine-tuned Artificial PSF (FAPSF) and the intensity of the related wings (i.e., Airy rings) are sensitive input parameters to the Lucy Richardson algorithm. As mentioned before, a list to verify the finding of FSS-stars is presented in Appendix \ref{appendix:checklist}.\newline
Using the positions from the NACO, SINFONI, NIRC2, and OSIRIS data analysis results in the Keplerian solution for the stellar candidate (named S4716) with an orbital period of 4 years (please find the R.A.-time and Dec.-time plot in Fig. \ref{fig:orbit}). Recently, GColl2022a observed in detail the periapse of $\rm \widetilde{S29}$ that coincides with the here presented Keplerian solution for S4716. We will comment on this confusion in the Discussion section. However, we include the positional GRAVITY data of 2021 for $\rm \widetilde{S29}$ as presented in GColl2022a and found a remarkably good match with the orbit of S4716 that is based on SINFONI and NACO observations covering the epochs 2003-2020.
\begin{figure}[htbp!]
	\centering
	\includegraphics[width=.5\textwidth]{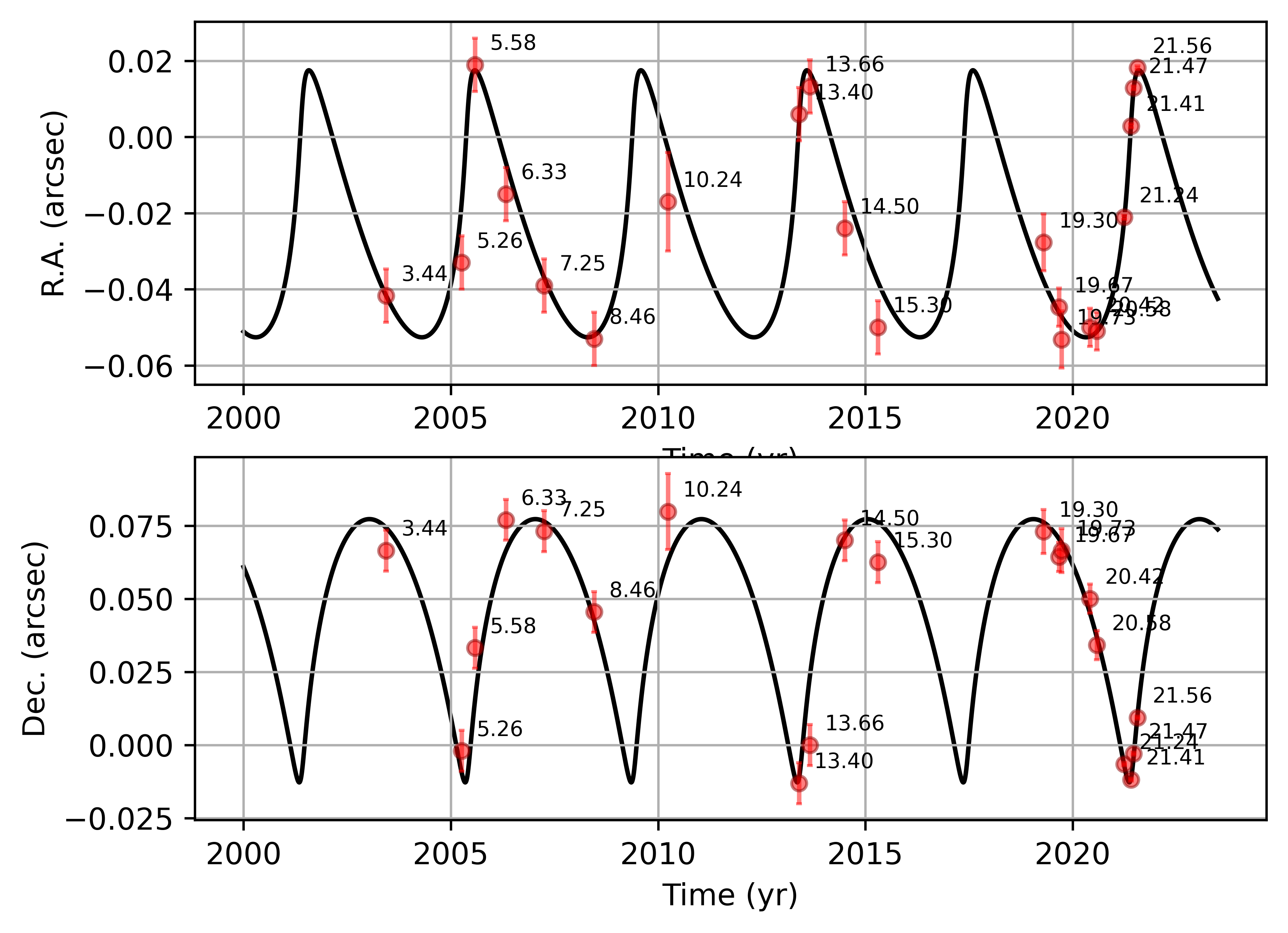}
	\caption{Orbit of S4716. Here we show the R.A. and Dec. component as a function of time. The epoch {\it t} is indicated by the numbers next to the data points. Because of the limited space, we choose the format "{\it t} - 2000" for the numerical values. As indicated in this figure, we identify S4716 in 16 epochs. In additon, we incorporate the observations by GColl2022a (see text for details). Blending events and the superposition with S2 can not be excluded and should be taken into account. For a detailed list of the data points, we refer to Table \ref{tab:positions_instruments_band}, Appendix \ref{sec:data_appendix}.}
\label{fig:orbit}
\end{figure}
In Table \ref{tab:orbit_elements}, we present the orbital parameters of S4716. We additionally list the orbital elements of S2 since this well-observed star is used to determine the position of Sgr~A* 
in agreement with the results described in \cite{Parsa2017}. The authors of \citet{Parsa2017} linked the infrared positional frame on a few milliarcsecond scale to the positions of known GC MASER stars obtained from the Very Large Array (VLA) radio observations.
The orbital elements listed for S2 in Table \ref{tab:orbit_elements} are taken from \cite{Do2019_S2} and agree reasonably well with the VLA results mentioned but also with GColl2021a.  
\begin{table*}[htb]
\centering
\begin{tabular}{ccccccc}\hline \hline
Source  & $a$ (mpc)&	$e$  &	$i$($^o$)&$\omega$($^o$)&	$\Omega$($^o$)&	$t_{\rm closest}$(yr)\\ \hline 
S2    &5.09 $\pm$ 0.01   &0.886 $\pm$  0.002 &133.49 $\pm$ ~0.40  &~65.89 $\pm$  0.75  &227.46 $\pm$~1.03  &2018.37 $\pm$ 0.02\\
S4716 (estimated)  &1.93 $\pm$  0.02  &0.756 $\pm$ 0.02 & 161.24 $\pm$ 2.80& 0.073 $\pm$  0.02&151.54 $\pm$ 1.54& 2017.41 $\pm$ 0.004\\
S4716 (max. likelihood)  &1.94 $\pm$  0.02  &0.756 $\pm$ 0.02 & 161.13 $\pm$ 2.80& 2.25 $\pm$  0.02&153.55 $\pm$ 1.54& 2017.41 $\pm$ 0.004\\
\hline \hline
\end{tabular}
\caption{Orbital elements of S2 and S4716. The estimated values for S4716 are based on a Keplerian fit model while the maximum likelihood stems from the MCMC simulation. See Appendix \ref{sec:mcmc} for a discussion of these values.}
\label{tab:orbit_elements}
\end{table*}
We find a reasonable agreement of the estimated Keplerian orbit with the maximum likelihood given by the Markov Chain Monte Carlo (MCMC) statistics (Table \ref{tab:orbit_elements}). While we limited the range of the pericenter passage to $\rm 2017.42\,\pm\,0.2$ years, we allowed every possibility for the mass and distance of Sgr~A*. 
We will elaborate on the statistical results for the mass and distance of Sgr~A* in Appendix \ref{sec:mass_distance_mcmc}.


\subsection{Parameters of S4716}
\label{sec:parameters_s4716}
\begin{figure}[htbp!]
	\centering
	\includegraphics[width=.5\textwidth]{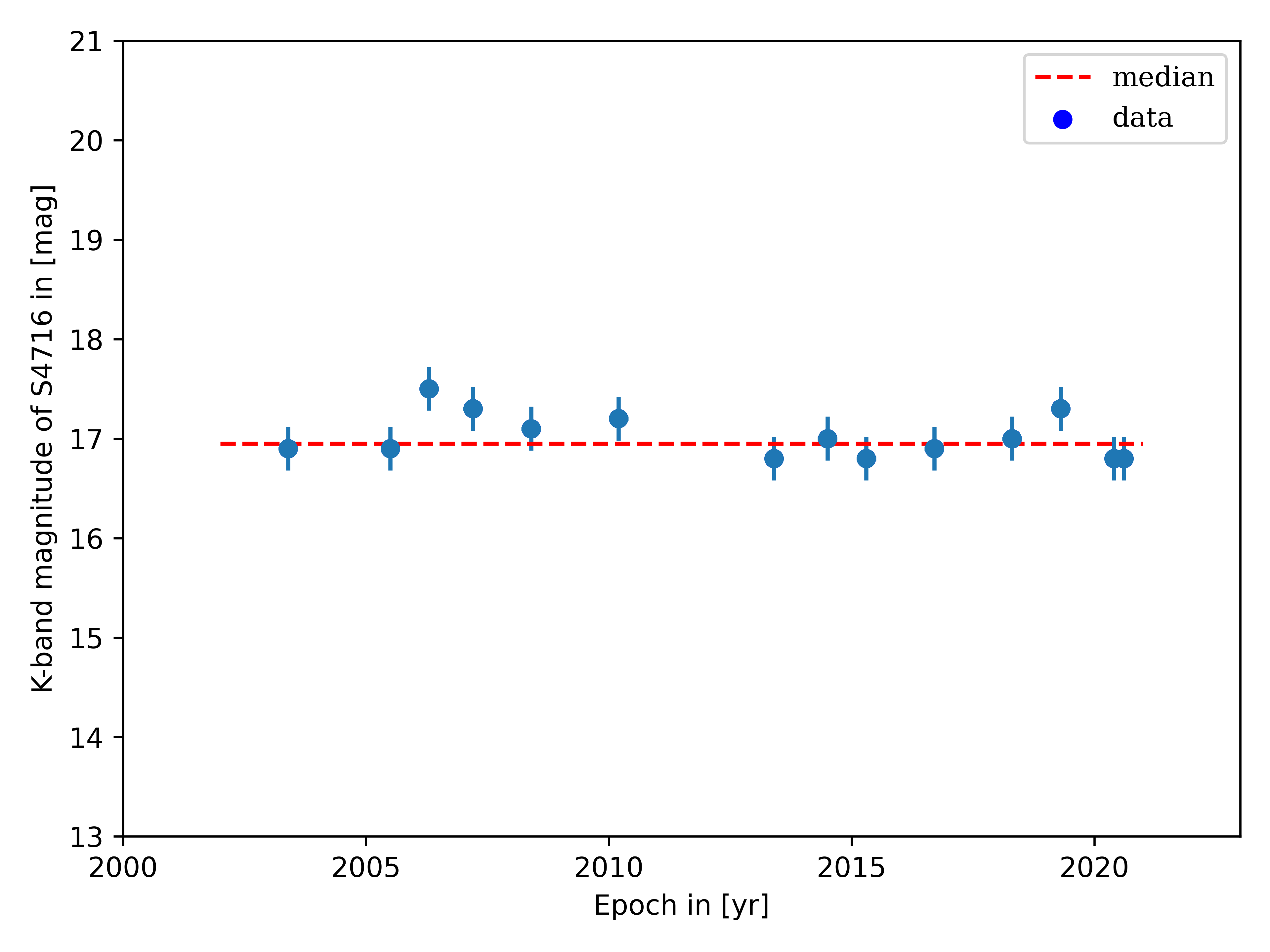}
	\caption{K-band magnitude of S4716 between 2003 and 2020. The red dashed line represents the first order polynomial fit. The blue points show the data. We derive an average K-band magnitude of $\rm 17.02\,\pm\,0.22$ mag. The magnitude is well inside the 1$\sigma$ uncertainty range.}
\label{fig:magnitude_s4716}
\end{figure}

The analyzed epochs in this work are showing S4716 as an isolated object. In these epochs, the identification resists the influence of the dominant PSF of S2. However, the variable footprint and distance of the S2 PSF affect the determination of the brightness of S4716. Therefore, we averaged the K-band magnitudes for individual identifications between 2003 and 2020 (see Table \ref{tab:positions_instruments_band}, Appendix \ref{sec:data_appendix}) to cover a large data baseline and counterbalance the mentioned fluctuations. Analyzing the data shown in Fig. \ref{fig:timeline}, we find an average K-band magnitude of $\rm 17.02\,\pm\,0.22$ mag (Fig. \ref{fig:magnitude_s4716}) by using
\begin{equation}
    \rm mag_{S4716}\,=\,-14.15\,+\,2.5\,\times\,log(counts_{S4716}/counts_{S2})
    \label{eq:magnitude}
\end{equation}
where S2 is used as the reference source \citep[][]{Schoedel2002}. Furthermore, we use
\begin{equation}
    f_{S4716}\,=\,f_{S2}\,\times\,10^{-0.4(mag_{S4716}\,-\,mag_{S2})}
\end{equation}
which is adapted from \cite{Sabha2012} to derive a K-band flux of $f_{S4716}\,=\,1.03\,\pm\,0.15$ mJy with $f_{S2}\,=\,14.72$ mJy and mag$\rm _{S2}\,=\,14.1$ \citep[][]{Schoedel2002}. Using the estimated magnitude, we follow the analysis of \cite{Peissker2020d} to calculate the mass of S4716 with
\begin{equation}\label{eq:mass}
 \log{(\frac{M_{S4716}}{M_{\odot}})}\,=\,k\cdot mag_{S4716}+b    
\end{equation}
where $k$ equals -0.1925 and $b=3.885$ (see Appendix \ref{sec:mass_estimator}). With Eq. \eqref{eq:mass}, we find a stellar mass for S4716 of $\rm M_{S4716}\,=\,4.04^{+2}_{-1}\,M_{\odot}$. Using the relation $\rm L\approx M_{S4716}^{\alpha}$ with $\alpha\,=\,3.5$, we estimate a stellar luminosity of $\rm log\frac{L_{S4716}}{L_{\odot}}\,=\,2.12^{+0.60}_{-0.41}$.\newline
With the semimajor axis of S4716 listed in Table \ref{tab:orbit_elements}, we find with Eq. \eqref{adv} a relativistic Schwarzschild precession of about $0.25^{\circ}\sim 15.1'$ and a periapse distance of $r_p\,=\,0.47-0.48$ mpc $\approx\,97-99$ AU. Furthermore, we estimate a relativistic parameter of $\Gamma\,=r_{\rm S}/r_{\rm p}\sim \,8.0-8.2\,\times\,10^{-4}$ which is comparable to that of S2 \citep[$6.8\,\times\,10^{-4}$, see][]{Peissker2020d}.\newline

\begin{figure}[htbp!]
	\centering
	\includegraphics[width=.5\textwidth]{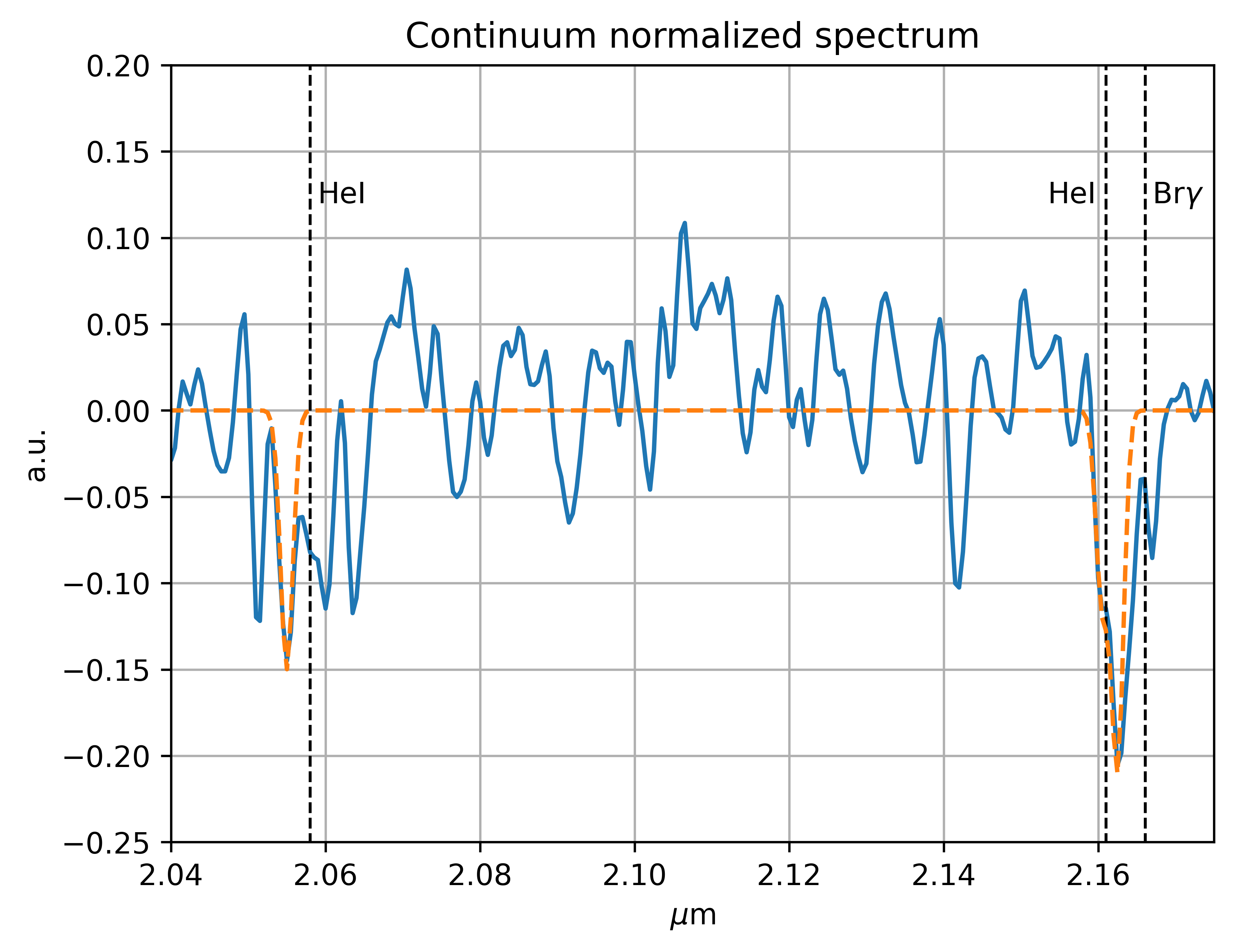}
	\caption{Spectrum of S4716 extracted from the SINFONI data cube of 2007. The rest wavelength of HeI $\rm 2p^1P^o-2s^1S$ at $2.058\mu m$ is indicated with a dashed line. At $2.161\mu m$ and $2.162\mu m$, the HeI doublet is located. The prominent absorption feature of the Br$\gamma$ line at $2.162\mu m$ is blue-shifted with respect to the rest wavelength at $2.166\mu m$. The redshifted spectrum of S4716 observed in 2009 is shown in Fig. \ref{fig:spec_4716_2009}, Appendix \ref{sec:spec_los}.}
\label{fig:spec}
\end{figure}

\begin{figure}[htbp!]
	\centering
	\includegraphics[width=.5\textwidth]{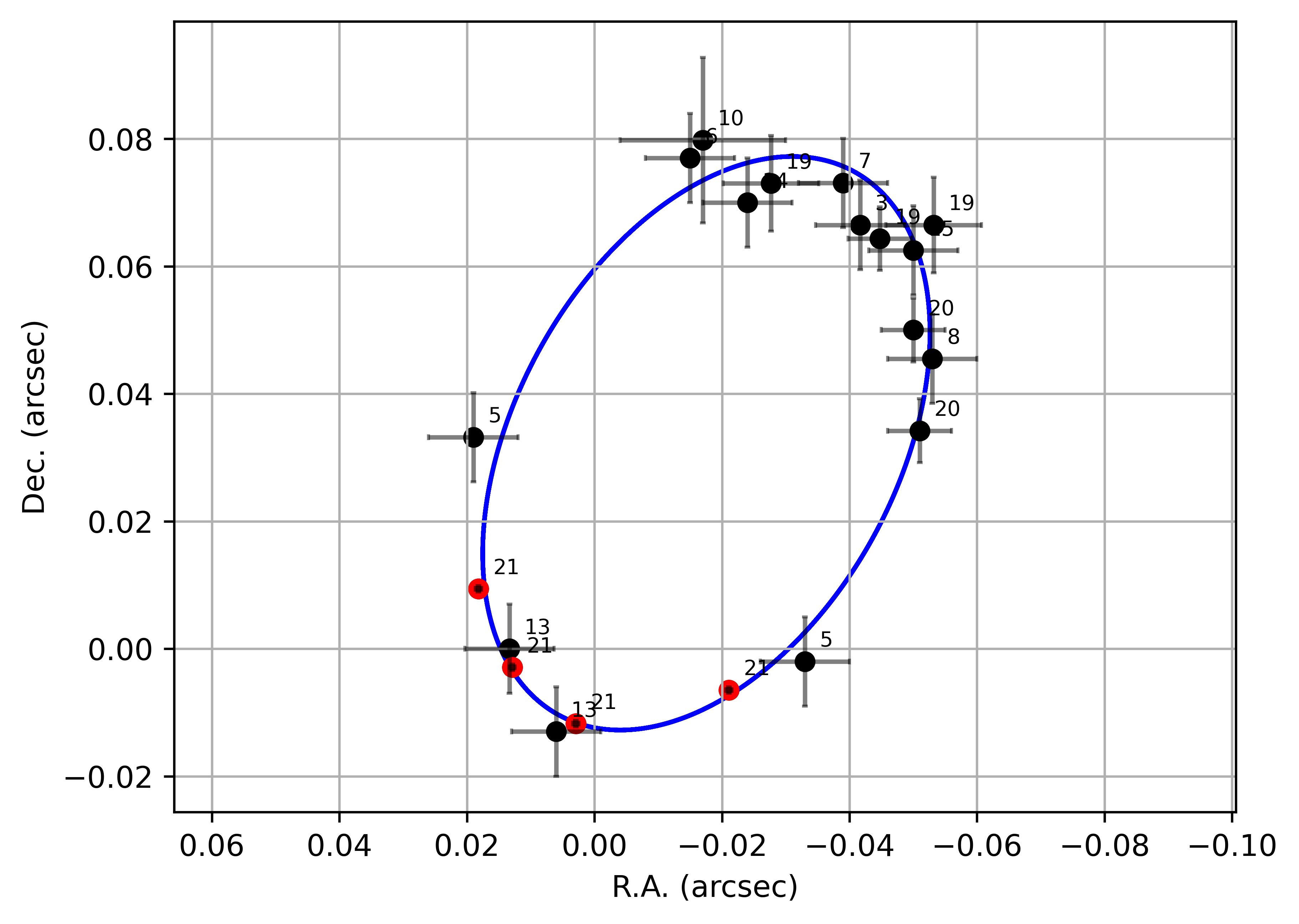}
	\caption{Projected on-sky view of the orbit of S4716. The red circled data points represent the GRAVITY measurements in 2021 are taken from \cite{GravityCollaboration2022a}. Please see Table \ref{tab:positions_instruments_band}, Appendix \ref{sec:data_appendix} for an overview of the shown data points.}
\label{fig:on_sky}
\end{figure}

Due to the isolated position of S4716 in 2007, we examine the related spectrum (Fig. \ref{fig:spec}) and find a Dopplershifted Br$\gamma$ emission line at $2.1625\,\mu m$. Using the Br$\gamma$ rest wavelength at $2.1661$, we derive an LOS velocity of about $v_{Br\gamma}\,=\,-430\pm 55$ km/s. In addition to the blue-shifted Br$\gamma$ line, we detect two HeI absorption lines. The helium line with the transition $\rm 2p^1P^o-2s^1S$ at a rest wavelength at $2.058\mu m$ shows the same blue-shift as the Br$\gamma$ line with an LOS velocity of $v_{HeI}\,=\,-437\pm 55$ km/s. The second detected HeI absorption line forms a shoulder feature \citep[][]{Habibi2017, Peissker2020d} that can be observed for many O/B-stars in the surrounding S-cluster. Since the spectrum of S4716 (Fig. \ref{fig:spec} and Fig. \ref{fig:spec_4716_2009}, Appendix \ref{sec:spec_los}) is comparable to S4711 \citep[][]{Peissker2020d}, we assume the same spectral type for both stars. Therefore, using the atlas of \cite{Hanson1996} and the approach of \cite{Peissker2020d}, we find similarities with a B8/9-V star. We follow up on this result in the following sections. 
This LOS velocity is consistent with the orbit fit result presented in Fig. \ref{fig:orbit}. The final LOS fit is shown in Fig. \ref{fig:los}, Appendix \ref{sec:spec_los}. Compared with the periapse LOS velocity of about 3000 km/s, the star is slow during its apoapsis, indicating the possibility of detecting S4711. The projected on-sky plot of the orbit and the implemented data points displayed in Fig. \ref{fig:on_sky} underline the probability of the observation of S4716 during apoapsis.  

Due to the high blue shift and width of the Br$\gamma$-line, the detection of the HeI doublet $7^3 F- 4^3 D$ and $7^1 F- 4^1 D$ is confused. Since the shape and velocity of the HeI lines at $2.161\,\mu m$ and $2.162\,\mu m$ are also influenced by the rotational speed $v\sin(i)$ and magnetic activity, we can only state a lower limit of the Doppler-shift of $\rm v_{HeI_{low}}\,=\,-210$ km/s. Because this estimate is influenced by the rotation of the star (and hence not necessarily completely correlated with $v_{Doppler}$), the lower value is in reasonable agreement with the former derived blue-shifted velocities of about $430$ km/s. Furthermore, the difference between the $7^1 F- 4^1 D$ line with the rest wavelength at $2.162\,\mu m$ and the spectral position of $v_{HeI_{low}}$ is $1.5\times 10^{-3}\mu m$. The difference is equal to $\rm 15\,\r{A}$. Now, we assume that the FWHM equals two times the Dopplershift difference between the line emission and the rest wavelength. This assumption is not entirely true due to the overall structure of the confused HeI doublet lines (Fig. \ref{fig:spec}). Therefore, we use the derived lower limit of $\rm v_{HeI_{low}}\,=\,-210$ km/s and the upper limit for the width of the line $\rm 2\times v_{HeI_{up}}\,=\,-420$ km/s to give a rough estimate of the stellar rotational velocity $\rm v_{rot}\,=\,v\sin(i)$ where we set $v\,=\,315$ km/s based on previous assumptions. However, this results in $v_{rot}\,=\,288\pm\,96$ km/s. 

\subsection{Stellar type}
\label{sec:type}
Here, we follow up on the stellar properties derived in the former subsection. We use the luminosity of $\rm log\frac{L_{S4716}}{L_{\odot}}\,=\,2.11^{+0.60}_{-0.41}$ and the mass with $\rm M_{S4716}\,=\,4.04^{+2}_{-1}\,M_{\odot}$ of S4716 as initial conditions for the BONNSAI (BONN Stellar Astrophysics Interface)\footnote{The BONNSAI web-service is available at \url{www.astro.uni-bonn.de/stars/bonnsai}.} \citep[][]{Schneider2014} analysis. The free available web-service BONNSAI is used to compare observed parameters with stellar models \citep[see][]{Brott2011}. As a boundary condition, the star analyzed should have a related mass between 3-100 M$_{\odot}$. The motivation of using the approach is the access to parameters that are not observed or to verify parameters, that suffer from increased confusion due to, for example, noise. However,we exclude $v_{rot}$ as a free parameter to investigate if BONNSAI finds a value that is comparable to the one estimated in the former subsection Sec. \ref{sec:parameters_s4716}. Although our number of priors is low, the resulting outcome is in good agreement with the input values (Table \ref{tab:bonnsai_results}).
\begin{table*}[htb]
\centering
\begin{tabular}{cccccc}\hline \hline
  & M in $\rm M_{\odot}$ &	$\rm log L/L_{\odot}$  & $\rm T_{eff}$ in K & R in $\rm R_{\odot}$ & $v_{rot}$ in km/s\\ \hline 
Input    & $4.04^{+2}_{-1}$ & $2.12^{+0.60}_{-0.41}$ & & &  \\
Output  & $3.20^{+0.62}_{-30}$  & $2.23^{+0.28}_{-0.20}$ & $12417.07^{+1787.69}_{-1860.28}$ & $2.45^{+1.34}_{-0.43}$ & $270.0^{+68.82}_{-257.48}$\\
\hline \hline
\end{tabular}
\caption{Results of the BONNSAI simulation of stellar evolution.}
\label{tab:bonnsai_results}
\end{table*}
We find an effective temperature of about $T_{eff}\,=\,12500$ K which underlines the classification of a B8/9-V star. Despite the wide uncertainty of the stellar rotational velocity, the outcome agrees with 1 $\sigma$ with our independently derived value in the previous section. The result of the stellar rotational velocity from the BONNSAI simulation marks an independent control point and confirms our assumptions in Sec. \ref{sec:parameters_s4716}.

\subsection{Mass and distance of Sgr~A*}
\label{sec:mass_distance}

Since S4716 with its short and small orbit offers a suitable choice of pinpointing to an exact value, we derive with the help of MCMC simulations the most likely distance and enclosed mass of Sgr~A*. For each plot, we use 10k iteration steps and open boundaries for the mass and distance ensuring a nonbiased outcome.\newline
The presented results in Fig. \ref{fig:mass_distance_1}-\ref{fig:mass_distance_3} (see Appendix \ref{sec:mass_distance_mcmc}) are used to estimate an averaged mass of $(4.023\,\pm\,0.087)\,\times\,10^6\,M_{\odot}$ for Sgr~A* where the uncertainty represents the mean absolute error.
Considering the distance of Sgr~A*, several attempts have been made to provide a consistent value. With the recent publications of \cite{Do2019_S2} and \cite{GravityCollaboration2022b}, a $4\%$ discrepancy is introduced regarding the distance of Sgr~A*. Here we incorporate both results because the averaged distance of the MCMC statistics for the S4716 orbit gives $(8.028\,\pm\,0.199)$ kpc or alternatively, $8.028$ kpc $\pm\,2\%$.

\subsection{Source identification in 2020}

\begin{figure*}[htbp!]
	\centering
	\includegraphics[width=1.\textwidth]{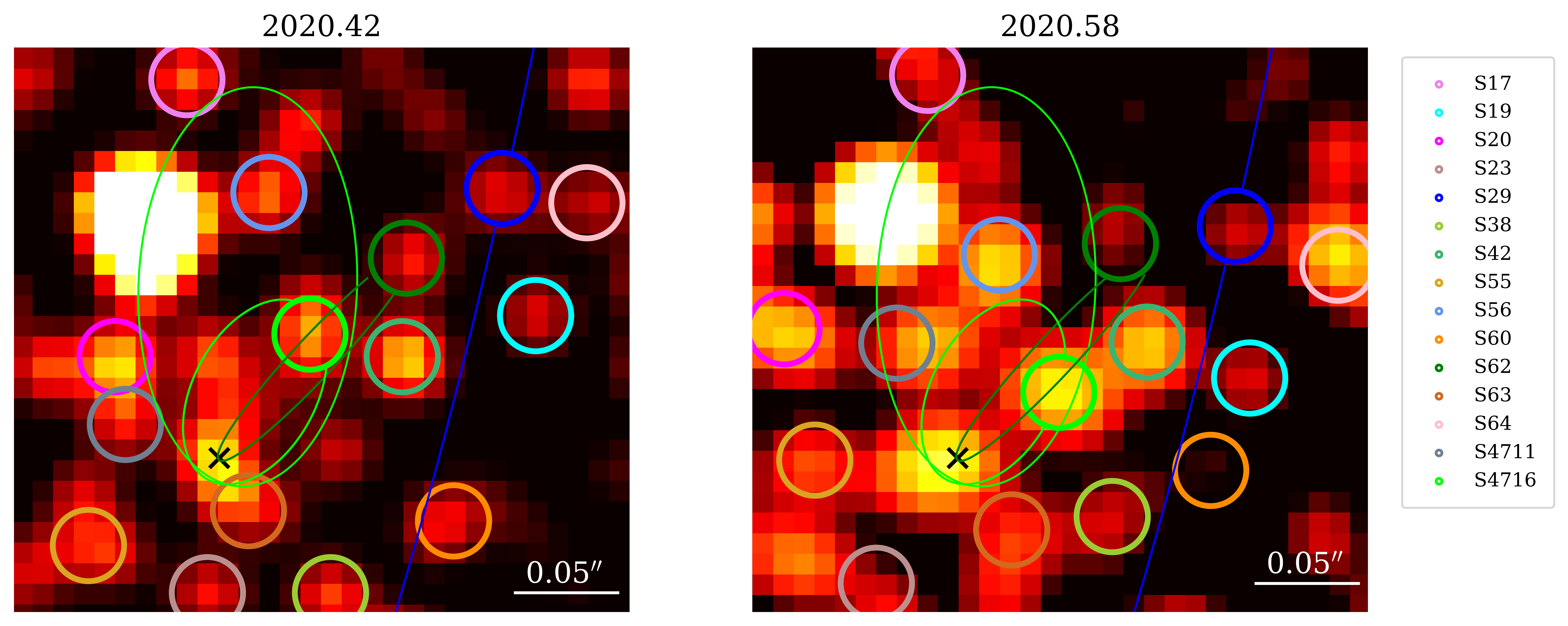}
	\caption{Observation of S4716 in 2020.42 and 2020.58 with OSIRIS. We highlight the detection of S4711 at the expected orbital position \citep[][]{Peissker2020d}. On these scales and sensitivity, a star like S300 are not visible (GColl2022a). However, S300 is part of the variable background and serves as an example for the challenging analysis of the imaging data. See the text for details. North is up, east is to the left.}
\label{fig:ident_2020}
\end{figure*}

As it is shown in \cite{Ali2020}, the number of S-stars with a related Keplerian orbital solution is much higher as discussed in \cite{Gillessen2009} and G17. We have shown in several publications, that the confusion in the S-cluster is enhanced because of nonidentified stellar objects \citep[see S148 in this work and Fig.2 in][]{peissker2021b}. With this degree of confusion, the observation of S62 or S4711 becomes challenging. However, gap-free long-term surveys of the vicinity of Sgr~A* are necessary to archive a uniform overview of the S-cluster. Since NACO and SINFONI were decommissioned in 2019, we used KECK data to continue the analysis of the S-stars with data that provides a complete overview of the cluster. In Fig. \ref{fig:ident_2020}, we show the data that was observed with the OSIRIS science cam with a spatial pixel scale of 10 mas. 
Consistent with the analysis covering 2003-2019, we find S4716 at the expected position in 2020.42 and 2020.58 (see also Fig. \ref{fig:orbit} for the related orbit position). As we follow the motion of S-stars \citep[][]{Ali2020}, we do find the listed traced stars at the expected positions. With the orbital elements for S29$\rm _{Gillessen}$/$\rm \widetilde{S29}$ given in \cite{Gillessen2009}, G17, and GColl2021a, we do not find a stellar source matching the high-precision measurements in the related publications.\newline 
However, it seems that S42 comes close to the orbital position predicated by the authors mentioned above. Due to the increased crowdedness in the vicinity of Sgr~A*, this star could be associated with S29 and would fit the orbital position derived by G17 using a 25-year data baseline. However, the derived pericenter passage for S29 is in 2025, excluding the possibility that GColl2022a observed this star during its closest approach in 2021.\newline
Additionally, we note the observation of S4711 as part of the FSS population in 2020 \citep[][]{Peissker2020d}. This stellar S-cluster member can be found at the expected and predicted orbital position (see Fig. \ref{fig:ident_2020}).


\subsection{Orbital properties of S4716 and spin of Sgr~A*}

The orbit of S4716 is the most compact orbit within the S cluster. With the semi-major axis of only $a=1.95\pm 0.02\,{\rm mpc}=402 \pm 4\,{\rm AU}$ (given by the maximum likelihood value), its orbital period is $P_{\rm orb}=4.02\pm 0.08$ years. The pericenter distance is $r_{\rm p}=a(1-e)=(99\pm 8)\,{\rm AU}=(2494 \pm 204)\,r_{\rm g}$ and the apocenter distance is $r_{\rm a}=a(1+e)=(706 \pm 11)\,{\rm AU}=(17783\pm 273)\,r_{\rm g}$, which makes it the first stellar orbit around the SMBH that is fully inside $1000\,{\rm AU}$, i.e. on the Solar-system length-scale. Given the eccentricity of $e=0.754 \pm 0.02$, the velocity increases significantly by a factor of $(1+e)/(1-e)\sim 7.13$ from the apocenter value of $(1116\pm 53)\,{\rm km\,s^{-1}}$ to the pericenter value of $(7956 \pm 381)\,{\rm km\,s^{-1}}$. In other words, when scaling to the light speed, the orbital velocity of S4716 increases from $(0.37\pm 0.02)\%$ of the light speed to $(2.65\pm 0.13)\%$ of the light speed, which is within $1\sigma$ uncertainty comparable to the S2 star at the pericenter. 

Using the definition of the relativistic parameter \citep{Parsa2017,Peissker2020d},
\begin{equation}
    \Gamma=\frac{r_{\rm S}}{r_{\rm p}}\,
\end{equation}
we calculate $\Gamma=(8.02 \pm 0.68)\times 10^{-4}$ for S4716, whose mean value is larger by a factor of $\sim 1.2$ and $\sim 1.4$ than for S2 and S4711 stars, respectively.

The relativistic quantities of interest -- the prograde apsidal (Schwarzschild) precession $\delta \phi$, gravitational redshift $v_{\rm gr}$, and the Lense-Thirring precession $\dot{\Omega}_{\rm LT}$ -- are essentially proportional to the first power of $\Gamma$ ($\delta \phi$, $v_{\rm gr}$) or its second power ($\dot{\Omega}_{\rm LT}$). Here we estimate these values as they could be of interest for future monitoring, e.g., with the Extremely Large Telescope. 

The prograde relativistic precession (Schwarzschild precesion) may be estimated as
\begin{equation}
    \delta \phi = \frac{6\pi GM}{c^2a(1-e^2)}=\frac{3\pi}{1+e} \Gamma\,,
\end{equation}
which for S4716 can be evaluated as $\delta \phi=(14.8 \pm 1.3)$ arcmin. The positional shift of the apocenter can be estimated as $\Delta s=r_{\rm a}\Delta \phi=3\pi a \Gamma=0.015\,{\rm mpc}\sim 0.4\,{\rm mas}$ that corresponds to less than $5\%$ of a pixel.

\begin{table}[]
    \centering
    \caption{Summary of relativistic parameters of S4716.}
    \begin{tabular}{c|c}
    \hline
    \hline
    Parameter  & Value  \\
    \hline
    Relativistic parameter $\Gamma$ &  $(8.02 \pm 0.68)\times 10^{-4}$\\
    Schwarzschild precession $\delta \phi$ [arcmin] & $14.8 \pm 1.3$ \\
    max. gravitational redshift [$\rm km\,s^{-1}$] & $120\pm 10$\\
    max. transverse Doppler shift [$\rm km\,s^{-1}$] & $106 \pm 10$ \\
    max. combined redshift [$\rm km\,s^{-1}$] & $226\pm 20$\\
    LT precession rate [$\chi=0.5$, arcsec\,yr$^{-1}$] & $1.11 \pm 0.13$\\
    \hline
    \end{tabular}
     \label{tab_rel_param}
\end{table}

Another general relativistic effect that affects the wavelength of the observed spectral lines is the gravitational redshift that for the pericenter may be evaluated as
\begin{equation}
    v_{\rm gr}=c\left[\left(1-\frac{r_{\rm S}}{r_{\rm p}} \right)^{-1/2}-1\right]\lesssim \frac{c\Gamma}{2}\,,
    \label{eq_gr_redshift}
\end{equation}
where the approximation on the right is valid for the pericenter. At the pericenter of S4716, the gravitational redshift amounts to $v_{\rm gr}=(120\pm 10)\,{\rm km\,s^{-1}}$. The gravitational redshift cannot be practically disentangled from the special-relativistic transverse Doppler shift, which for S4716 amounts to $v_{\rm t}=c[(1-(v_{\rm p}/c)^2)^{-1/2}-1]=(106 \pm 10)\,{\rm km\,s^{-1}}$ at the pericenter. The combined maximum relativistic redshift at the pericenter for S4716 then is $(226\pm 20)\,{\rm km\,s^{-1}}$. 

Finally, the compact orbit of S4716 is affected by the frame-dragging or the Lense-Thirring (LT) effect in the spacetime of a rotating (Kerr) black hole. The orbit-averaged change of the angular momentum direction can be calculated with
\begin{equation}
    \left(\frac{\mathrm{d}\mathbf{L}}{\mathrm{d}t}\right)=\frac{2G^2M^2}{c^3(1-e^2)^{3/2}a^3}\mathbf{\chi}\times \mathbf{L}\,,
\end{equation}
where $\mathbf{\chi}$ is the direction of the Sgr~A* spin and $\mathbf{L}$ is the angular momentum direction of S4716. The frequency of the angular momentum precession can be evaluated as
\begin{equation}
    \nu_{\rm L}=\frac{2G^2M^2\chi}{c^3a^3(1-e^2)^{3/2}}=\dot{\Omega}_{\rm LT}\,,
    \label{eq_nodal_precession}
\end{equation}
which is equal to the orbit-averaged precession frequency of the nodal axis. The expression given by Eq.~\eqref{eq_nodal_precession} can be rewritten as \citep{Peissker2020d},
\begin{equation}
    \nu_{\rm L}=\dot{\Omega}_{\rm LT}=\Gamma^2 \frac{\chi c}{2r_{\rm a}}\sqrt{\frac{1-e}{1+e}}\,.
\end{equation}
The Lense-Thirring precession rate of S4716 for the Sgr~A* spin of $\chi=0.5$ is $\nu_{\rm L}=1.11 \pm 0.13$ arcsec per year, which is larger than for S2 and S4711, see \citet{Peissker2020d} for the comparison in terms of the FSS-population. 

Since S4716 appears to have one of the largest reliable Lense-Thirring precession rates (S62 and S4714 have large uncertainties), one can in principle turn the LT rate around to constrain the spin of Sgr~A*. \citet{Ali2020} introduced the disk-like kinematic structure of the S cluster in 3D space. Based on that, \citet{Fragione2020} constrain the spin of Sgr~A* to $\chi\lesssim 0.1$ using the argument that the Lense-Thirring precession would disrupt the disk structure \citep[see also][]{Fragione2022}. S4716 seems to deviate from the disk structure with its inclination of $\sim 161^{\circ}$ being offset by more than $70^{\circ}$ from the disk inclination peak close to $90^{\circ}$. This could be caused by the vector resonant relaxation (VRR), which is the fastest for the inner orbits of the S cluster. The timescale of the VRR behaves with the semi-major axis approximately as $T_{\rm VRR}\propto a^{\gamma/2}$ for the power-law slope $\gamma$ of the density cusp of late-type stars, $n_{\star}\propto r^{-\gamma}$. Therefore, the inclination distribution of S stars is expected to become wider towards smaller semi-major axes. S4716 has the maximum-likelihood longitude of the ascending node of $\Omega_{S4716}\sim 150^{\circ}$. Using this value, S4716 falls into the $\Omega$ distribution of the ``black'' disk with the mean value close to $180^{\circ}$. Under the assumption that the temporal evolution of $\Omega_{S4716}$ is driven mainly by the Lense-Thirring precession, one can put the constraint on the $\dot{\Omega}_{\rm LT}$ requiring that the Lense-Thirring timescale is comparable to or larger than the lifetime of S stars in order not to completely obliterate the non-isotropic $\Omega$ distribution of S stars. Using the expected deviation of $\Omega_{S4716}$ from the mean value for the ``black'' disk, $\Delta \Omega=\overline{\Omega}-\Omega_{S4716}\approx 30^{\circ}$, we can write,
\begin{equation}
    \tau_{\rm LT}=\frac{\Delta \Omega}{\dot{\Omega}_{\rm LT}}\gtrsim \tau_{\rm life}\sim 6.6\times 10^6 \rm{yr}\,.
\end{equation}
The required precession rate is quite small, $\dot{\Omega}_{LT}\lesssim 0.0164$ arcsec per year, which constrains the spin of Sgr~A* to $\chi\lesssim 0.0074$, i.e. consistent with $\chi<0.1$ found by \citet{Fragione2020}. The requirement of the well-ordered, non-isotropic 3D structure of the S cluster, i.e. consisting of multiple well-defined coherent disks in 3D, puts quite a stringent constraint on the Sgr~A* spin towards small values. Even if the compact orbit of S4716 has already made one turn of $\Omega$ by $\pi$ radians during its lifetime, i.e., it would be now located where it was initially within one of the disks, the precession rate would still be small $\dot{\Omega}_{\rm LT}\lesssim 0.115$ arcsec per year and the SMBH spin at most $\chi\lesssim 0.052$.

\section{Discussion}
\label{sec:discussion}
In this section, we categorize our findings and discuss possible impacts on the detection of the compact and extended mass in the center of our galaxy. After the pericenter passage of S2, we reidentified S4711 in 2020 with OSIRIS that were previously investigated in \cite{Peissker2020d} and introduce a new S-star called S148. We would like to highlight a blend star event in 2017 of S148 and S4716 that is discussed in detail in Appendix \ref{sec:s148}.

\subsection{The observation of S4716 between 2003 and 2021}

While we continue to analyze the S-stars and the immediate environment, we recognized an unidentified stellar source in 2019 using public NIRC2 and NACO data. With our orbital tracking list, we identified the surrounding stars of this new stellar source neglecting an identification problem. Because of the proper motion that is directed from east to west with a prominent curvature in 2019.30, 2019.36, 2019.49, and 2019.67, we were able to identify this new star in 2020.42 and 2020.58 following a traceable trajectory. As GColl2022a claims to trace S29 by promoting a new orbital solution compared to \cite{Gillessen2009}, G17, GColl2021a, and \cite{peissker2021b}, we provide an alternative interpretation of the interferometric GRAVITY data observed in 2021 with robust imaging data between 2003 and 2020.\newline 
By investigating published and archival data, we identified this new source that we call S4716 in 16 epochs between 2003 and 2020 (see Table \ref{tab:positions_instruments_band}, Appendix \ref{sec:data_appendix}). Since crowding is a general problem in the S-cluster, S4716 is confused with S62 in 2004, 2011, and 2012. Moreover, the close distance of S2 hinders a confusion-free detection of the star in 2016 (Appendix \ref{sec:s4716_2016_2017_2018}). 
We underline the robustness of our finding by a photometric analysis covering data between 2003 and 2020. In agreement with the 1$\sigma$ uncertainty range, we find an averaged K-band magnitude for S4716 of $\rm mag_K\,=\,17.02\pm 0.22$ mag. Although fluctuations could be explained by a variable background, the extinction toward the Galactic center should also be considered \citep[][]{Fritz2011}. Furthermore, \cite{Schoedel2010} reported a median extinction of $A_K\,=\,2.74 \pm 0.3$ mag which is supported by the findings of \cite{Peissker2020c}. Although it is plausible that the infrared magnitude of the general S-cluster population can suffer from extinction variations, it is rather unlikely that these fluctuations show an imprint on the small S4716 orbit. The overall on-sky coverage of the S4716 orbit is in the order of around 0.6$\%$ compared to the S-cluster size. 
If the variation of the extinction causes the uncertainties observed for S4716, they are very well covered by our derived standard deviation of 0.22 mag \citep[for comparable values, see][]{Schoedel2010, Fritz2011, Gautam2019}. We note that the faintest K-band magnitude values (Fig. \ref{fig:magnitude_s4716}) are observed with NIRC2/KECK implying a general characteristic of the detector. Although the wavelength coverage of the investigated data is comparable (see Table \ref{tab:positions_instruments_band}), the faint NIRC2 data points could imply a different detector sensitivity. Here, one has to take into account that the on-source integration time of the NACO data is increased compared to the KECK data (see Appendix \ref{sec:data_appendix}) resulting in a presumably higher S/N ratio. Because faint sources like S4716 are sensitive to background emission, the different on-source integration time is a suitable explanation for the observed effect. We also note that we estimate the magnitude for S4716 with the ratio indicated in Eq. \ref{eq:magnitude} which should be independent of the instrument.\newline
Furthermore, we witnessed a blend star event with a new identified S-star called S148 in 2017 (see Appendix \ref{sec:s148}). In 2018, we reidentify S4716 at the expected projected on-sky position on its orbit. The data of 2016-2018 is shown in Appendix \ref{sec:s4716_2016_2017_2018} where we provide photometric information of S148.\newline
For the epochs 2007 and 2009, we extract the spectrum of S4716 using SINFONI data. The line-of-sight velocity of about -500 km/s and +1690 km/s matches the orbital fit model providing an independent confirmation of the detection. It is generally plausible that the chance of observing this star during the apoapse is increased due to the slow proper motion during this orbital section.\newline 
In addition to the observation of S4716, we find S4711 at the expected position during its apoapse in 2020 using the orbital elements from \cite{Peissker2020d}. Likewise for S4716, the star S4711 can be easier observed during the apoapse.

\subsection{Systematics}

Due to the number of different intruments, the analysis is sensitive to systematic uncertainties. As we pointed out in Sec. \ref{sec:methodology}, the data from different sources seem to result in agreeing high precission outcomes. However, \cite{Plewa2015, Plewa2018} analyze the effects of the NACO detector on the data observed in GC. Plewa et al. furthermore discuss faint unrecognized objects that might influence the background emission and, therefore, the source determination. In, e.g., \cite{Peissker2020d}, we faced these problems by identifying several new S-cluster members. Without a doubt, the S-cluster has many more sources as is known today \citep[for a review, see][]{Clenet2019}. In addition to the analysis of the NACO detector and its influence on astrometric measurements, \cite{Jia2019} investigates positional uncertainties as a function of the magnitude of the K-band as observed with KECK. For a 17 mag faint star like S4716, the positional uncertainty is in the range of 0.1-1 mas. However, Jia et al. also show that uncertainties up to 10 mas are possible in the 17 mag K-band range.\newline
We can directly compare the position of S2/S0-2 with the orbital fit provided by \cite{Do2019_S2} and GColl2022a. Furthermore, we choose S29 as a proper choice to investigate the dimensions of the systematics. This star was observed between 2002 and 2019 using mainly NACO data\footnote{Except for 2014 and 2019 where we used SINFONI data.} \citep[][]{peissker2021b}. In Fig. \ref{fig:ident_2020}, we find S29 in the expected position in two different epochs. We can conclude that the systematics and different spatial resolutions are not the dominant origin of uncertainties.

\subsubsection{The proper motion of $\widetilde{S29}$}

\begin{figure}[htbp!]
	\centering
	\includegraphics[width=.5\textwidth]{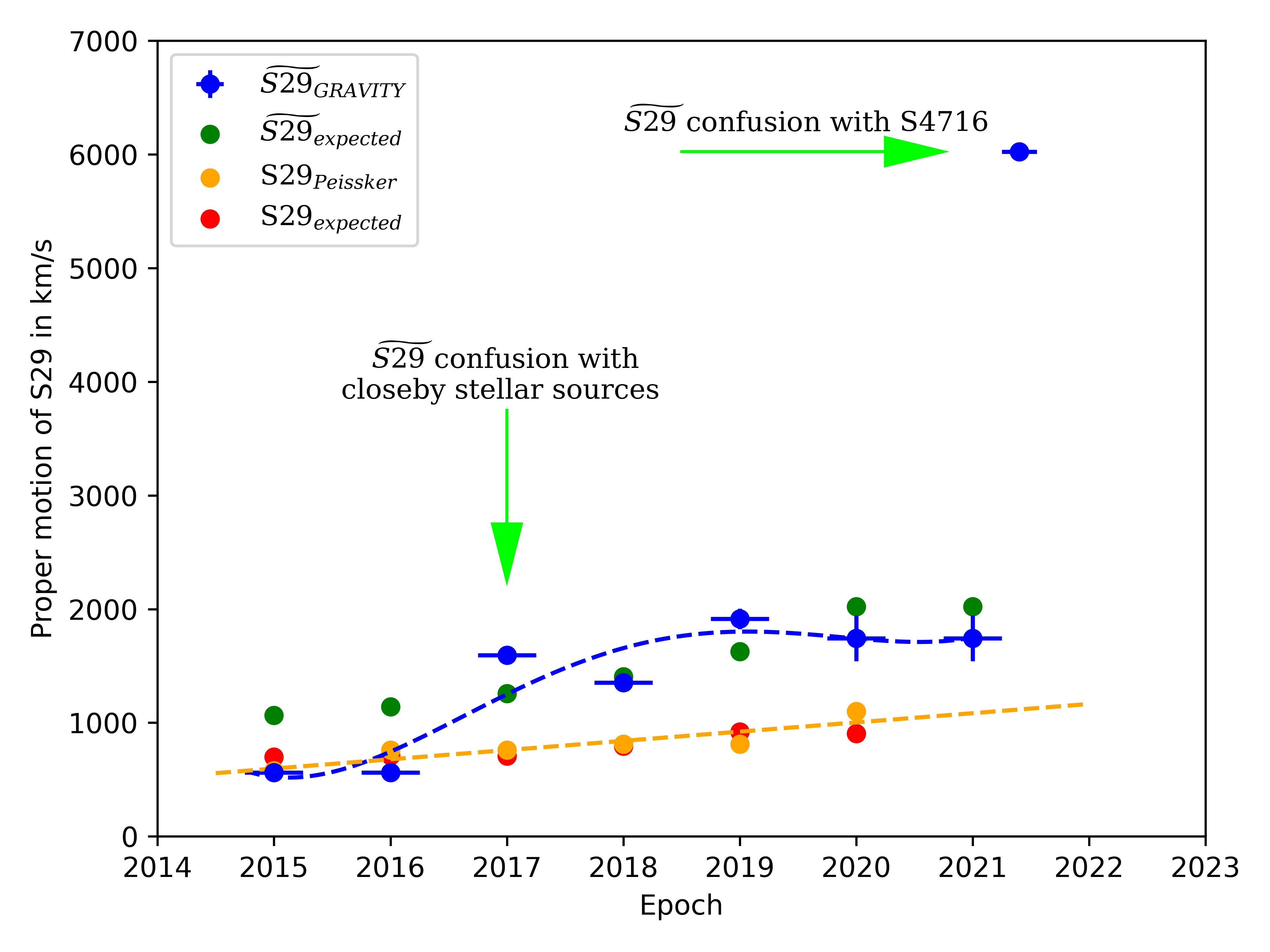}
	\caption{Proper motion of $\rm \widetilde{S29}$ as presented in GColl2022a. Because of the missing listening of the exact epoch per year in the related publication, we used a large x and y uncertainty of $\pm\,0.5$ years and $\pm\,200$ km/s. The red curve represents a 3rd order polynomial. Inspecting the proper motion underlines a identification problem for $\rm \widetilde{S29}$.}
\label{fig:pm_s29}
\end{figure}

While we are not questioning the high precision measurements of GColl2022a, we would like to discuss a possible confusion in the presented data of the related publication. We remind the reader about the nomenclature that is introduced in the introduction of this work: we distinguish between S29 \citep[][]{peissker2021b}, S29$\rm _{Gillessen}$ (G17), and $\rm \widetilde{S29}$ (GColl2022a). While the first two listed publications provide different orbital solutions to the trajectory of S29 based on an almost 20-30 year data baseline, the common consensus is a periapse passage that does not match $\rm \widetilde{S29}$ as proposed by GColl2022a. Hence, we raise the question which star the authors actually have observed since the mentioned data baseline of almost 30 years analyzed in G17 covers a significant part of the S29$\rm _{Gillessen}$ orbit.\newline
Unfortunately, the authors of GColl2022a missed the chance of indicating the precise epoch for the presented $\rm \widetilde{S29}$ orbit. Since the S-stars are characterized by a high proper motion \citep[][]{Genzel2000}, this will have an impact on the uncertainty of the here presented discussion. Anyhow, by analyzing Figure D.1. of GColl2022a, we extract x- and y- distances\footnote{We use the software Digitizer (Thomas Ott, MPE Garching) for the extraction of data points.} of S29 with respect to Sgr~A* which are listed in Table\ref{tab:pm_s29}.
\begin{table}[htb]
\centering
\begin{tabular}{cccc}\hline \hline
  Year & $\Delta$x in [mas] &	$\Delta$y in [mas]  & v$\rm _{prop}$ in  [km/s] \\ \hline 
2015    & 148.1 & 237.0 & -  \\
2016    & 140.7 & 222.2 & 562.1 \\
2017    & 125.9 & 177.7 & 1593.3  \\
2018    & 111.1 & 140.7 & 1353.9  \\
2019    &  88.9 &  88.9 & 1914.7  \\
2020    & - & - &  1742.63 \\
2021    &  26.6 &   7.4 & 1742.63  \\
\hline \hline
\end{tabular}
\caption{Proper motion of $\rm \widetilde{S29}$ derived from GColl2022a (see their Fig. D.1.). The here listed proper motion is calculated by the year-to-year difference. The authors of the related publication do not provide a data point for 2020. Hence, we use the positional difference between 2019 and 2021. It is implied, that the proper motion of S29 is at least in 2020 or 2021 too low.}
\label{tab:pm_s29}
\end{table}
We use the values from Table \ref{tab:pm_s29} to derive the proper motion between two consecutive years by calculating the positional change over 365 days. We are aware that this is a rough approximation, but it should serve the purpose since GC observations with NACO can not be executed throughout the whole year\footnote{From Paranal/Chile, the GC is only observable between end of February up until the beginning of October.}. For the calculation of the associated proper motion, we include the GC distance relation $\rm 14\,mas\,=\,100\,AU\,=\,150\cdot 10^8\,km$. With this, we calculate the related proper motion given in Table \ref{tab:pm_s29}. For the data in 2020, the authors of GColl2022a are not providing a data point. Therefore, we calculate the difference between 2019 and 2021 and divide the resulting value by 2. For the 4 months in 2021 during the pericenter passage of the star observed with GRAVITY, we calculate a proper motion of 6025 km/s. However, fitting a 3rd order polynomial to the data underlines the confusion regarding S29, $\rm \widetilde{S29}$, and S4716 (Fig. \ref{fig:pm_s29}).
As mentioned in GColl2022a, the star S29 is confused between 2015 and 2017. We indicate in Fig. \ref{fig:pm_s29}, that the disagreement with the orbital solution and the derived proper motion starts in 2017. We furthermore find that the star decelerates in 2018, gets faster in 2019, and decelerates either in 2020 or 2021. We can conclude that the velocity of the observed star in 2021.5 by GColl2022a exceeds the proper motion of $\rm \widetilde{S29}$ that is derived from the presented imaging data in the associated publication. 

\subsection{Extrapolating high-precission data}

\begin{figure}[htbp!]
	\centering
	\includegraphics[width=.5\textwidth]{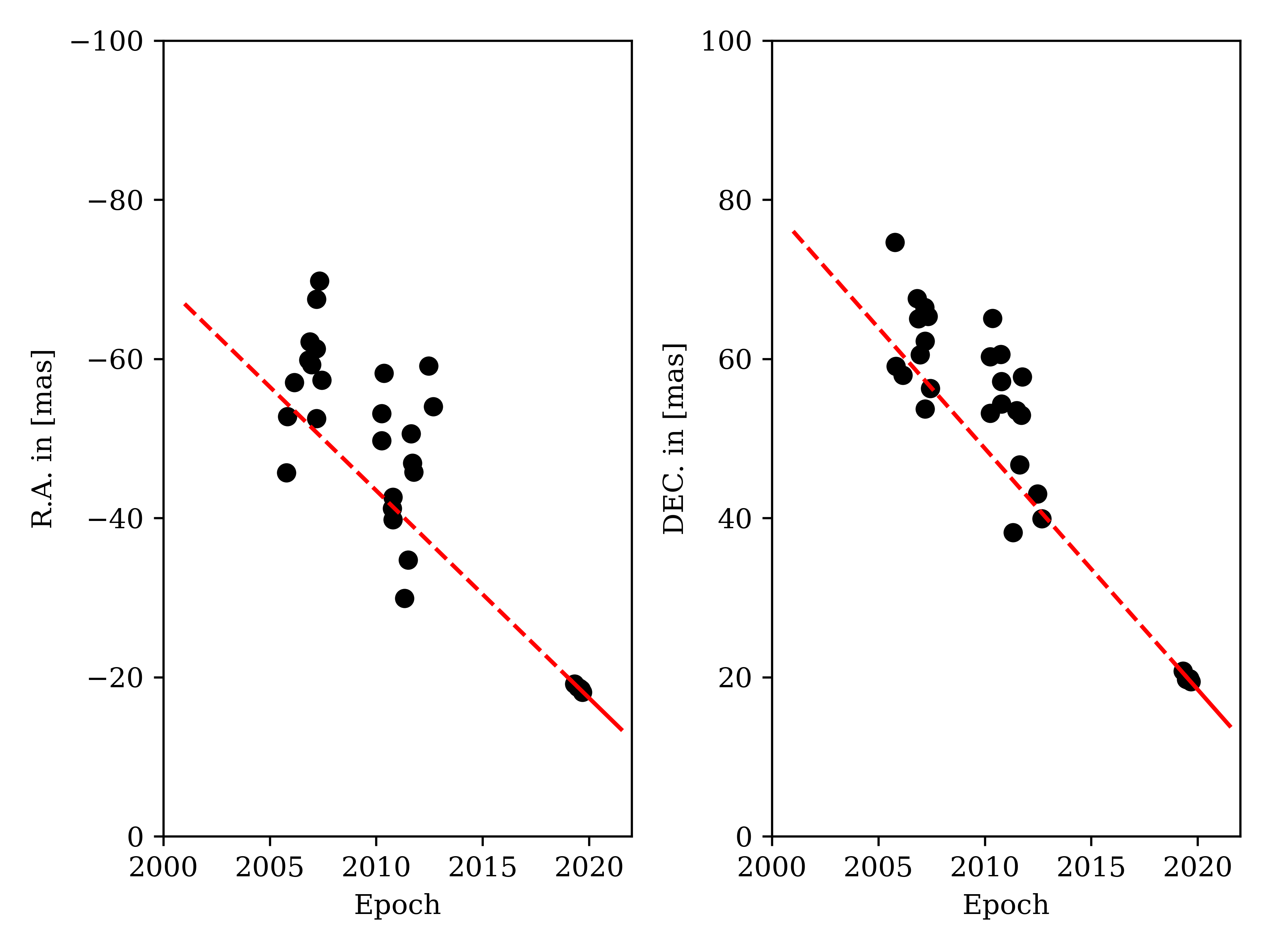}
	\caption{Linear motion of S62 as proposed by \cite{Gillessen2009}, \cite{Gillessen2017}, and GColl2022a. Data before 2015 was observed with NACO, after 2015 with GRAVITY. The non-linear motion of S62 is obvious for the data points before 2015 and follows the trend that we already presented in \cite{Peissker2020a}. The large gap between the NACO- and GRAVITY-data spans over 7 years. This data is extracted from GColl2021a.}
\label{fig:pm_s62_linear}
\end{figure}

We would like to emphasize that we are not questioning the capabilities of GRAVITY. However, using high-precision data and extrapolating these data to a crowded cluster may result in misidentifications. To underline this point, we shortly want to discuss the observation of the linear motion of S62$_{\rm linear}$ \citep[see][]{Gillessen2009, Gillessen2017}. To date, it is not clear how a star can be gravitational bound to Sgr~A* without showing any signs of acceleration or curvature. We discussed this already in \cite{peissker2021b} but we want to visualize the basic problem that emerges from this approach. For this, Fig. \ref{fig:pm_s62_linear} shows the extracted data points from GColl2021a. 
Data prior to 2015 observed with NACO suggest a non-linear trajectory for S62$_{\rm linear}$. Although we are aware of the noisy character of the NACO and SINFONI data, a non-linear trend for the data points is clearly observed before 2015. After 2015, the data observed with GRAVITY were extrapolated, resulting in the linear classification of S62. We point out that S55 was classified as a linear star \citep{Gillessen2009} before \cite{Meyer2012_s55} derived a Keplerian orbit with an orbital period of 11.6 years. The misindetification of S55 and S62 underlines our argument that extrapolating high-precision data into a crowded cluster may result in the confusion of stars. This confusion is reflected by the discontinuous velocity of $\rm \widetilde{S29}$ as shown in Fig. \ref{fig:pm_s29}.

\subsection{Limitations of the analysis of the S-cluster using S300}

Inspecting the data sets reveals stars with a variable magnitude and high confusion. While these problems are already discussed in \cite{Sabha2012} and \cite{Peissker2020a}, we would like to discuss some of the challenges with the robust observation of S62, S4711, and S4716 (see Fig. \ref{fig:ident_2020}) over two decades. While all three listed stars can be observed in the majority of available data sets, S2 hinders the observation of S62, S4711, and S4716 in the years during its periapse passage. As we know from the observation of S300 in GColl2022a, the diffuse background of the here presented imaging data can suffer from variations that are challenging to determine \citep[][]{Sabha2012, Eckart2013}. For S300, the range of K-band magnitudes are between 19.3-20.5 mag. Based on the modeling of \cite{Jaroszynski1999}, it is plausible that other stars like S300 close to Sgr~A* exist that could interfere with the confusion-free detection of S62, S4711, and S4716. Donating such a star with S300*, we assume a K-band magnitude of $19.9\,\pm\,0.6$ mag. Using S2 as the reference star with a K-band magnitude of 14.15 mag, the (pixel-)counts from an observation of S300* are $0.5\%$ of the peak maximum. Since S4711-S4715 are indicating the detection limit for the here discussed imaging data, it is reasonable to subtract up to $2\%$ of the image flux (i.e. background subtraction). With this approach, the image information is preserved since the flux of every observed source is above the upper limit of $2\%$. However, the $2\%$ limit also shows that stars like S300* do exist inside the faint background and could presumably have an measurable impact on the extended mass. It is important to note that this $2\%$ limit shows a different impact on the image area: the background close to Sgr~A* is higher compared to the outer region of the S-cluster \citep[see Fig. \ref{fig:bs} and][]{Sabha2012}. Hence, the chance of losing information especially close to the image edges is increased with the here discussed methods.     

\subsection{The compact and extended mass}

From the MCMC statistics, we derive an enclosed mass $m_{enc}$ of $(4.023\,\pm\,0.087)\,\times\,10^6 M_{\odot}$ for the orbit of S4716 which numerically agrees with the results of \cite{Boehle2016}. This derived result includes the compact mass $m_{com}$ that is associated with Sgr~A* and furthermore the extended mass $m_{ext}$. Here, $m_{ext}$ is a measure for the dark mass which should have a measurable imprint on the orbit of S4716. Deviations from the Keplerian elliptical orbit can arise either due to an extended mass inside a stellar orbit (retrograde periapse precession) or due to post-Newtonian effects of the curved spacetime around a point mass \citep[prograde Schwarzschild precession, see Eq.~\eqref{adv};][]{Weinberg1972, Rubilar2001, Mouawad2005}.
 In reality, both Newtonian and relativistic effects are expected to contribute to a smaller or greater extent.

Following this approach, we define the measured enclosed mass as a sum of the compact and extended mass with
\begin{equation}\label{enclosed_mass}
    m_{enc}\,=\,m_{ext}\,+\,m_{com}\,.
\end{equation}
When we use Eq. \eqref{enclosed_mass} in units of $m_{enc}$, we can write $1\,=\,m_{ext}/m_{enc} \,+\,m_{com}/m_{enc}$ where as a consequence $m_{com}=m_{enc}$ yields $m_{ext}=0$ for a pure point-mass model, i.e. only the SMBH would contribute to the enclosed mass. For this, we assume a sphere with a volume $V_{sphere}$ that includes both mass components.\newline
In the following, we assume that part of the mass located within the S4716 orbit is due to an extended mass $m_{ext}$ component surrounding the otherwise heavy compact mass $m_{com}$ of SgrA*. The amount of an extended mass can be estimated by putting a limit on the corresponding retrograde Newtonian periastron shift, as long as it is large with respect to the prograde relativistic shift.
The orbit of S4716 allows us to make a statement on the amount of $m_{ext}$ contained in the volume that is transversed by the S4716 star 
on its orbit. The distribution of any extended mass within $V_{sphere}$ is unknown. However, if $m_{ext}$ is more concentrated towards the SMBH or towards the rim of $V_{sphere}$, then the length of the orbital sections within that extended mass and correspondingly the amount of the Newtonian retrograde shift is diminished. 
Hence, assuming the first-order homogeneous mass distribution puts a realistic upper limit on the amount of $m_{ext}$. \cite{Rubilar2001} as well as \cite{Mouawad2005} have investigated the effect of an extended mass on stellar orbits using the formalism provided by \cite{Jiang1985} in combination with orbit calculations using the post-Newtonian approximation of General Relativity \citep[][]{Weinberg1972}. For S4716, we will derive in the following a simple estimate of the order of magnitude of the extended mass within the $V_{sphere}$ that is transversed along its orbit.\newline
For a body travelling on an elliptical orbit, the orbital time scale $T$, the semi-major axis $a$, and the central mass $M$ are related via the third Kepler law
\begin{equation}
a^3 = \frac{GMT^2}{4\pi^2}~~~.
\end{equation}
Here, $G$ is the gravitational constant. Furthermore, one finds for an elliptical orbit
\begin{equation}
b=a \sqrt{1-e^2}~~~.
\end{equation}
Here, $b$ is the semi-minor axis and $e$ the eccentricity of the orbit. If the amount of extended mass is $m_{ext}$, then varying the Mass 
from $M$ to $M+m_{ext}$ results in a change of orbital elements. Requiring that $a$ for each semi-elliptical section of the 
rosetta-shaped Newtonian orbit remains constant, the changes are restricted to $e$ and $b$, correspondingly. Hence, for successive
semi-elliptical sections, the argument of periapsis $\omega$ changes in a retrograde way. In the fortunate case of S4716, the orbit is almost face on, therefore the uncertainty of $\delta\omega$ of the argument of periapsis $\omega$ can immediately be translated to an uncertainty of the compactness of the central mass $M$ due to an additional extended mass component $m_{ext}$.\newline 
The MCMC fit of the  S4716 orbit gives a 1$\sigma$ uncertainty of $\delta\omega=0.023^\circ$ (0.0004 rad).
This corresponds to a possible extended mass contribution of 7$\times$10$^{3}\,M_{\odot}$ on the 1$\sigma$ level and a 2$\times$10$^{4}\,M_{\odot}$ contribution on the 3$\sigma$ level. This is more than 25 times smaller than the mass uncertainty given by the MCMC fit of the S4716 orbit.
Our derived estimate of the extended mass is therefore on the same level as the robust estimate by \cite{Mouawad2005} for the 2.5 times larger S2 orbit and the corresponding limits summarized by \cite{Gillessen2009} and G17. 

With $m_{ext} \le 2\times10^4$ \solm ~~and $m_{enc} \sim 4\times10^6$ \solm ~we
can then directly derive
a limit on the entire compact mass component associated with SgrA* as
 $m_{com} \lesssim 3.98\times10^6$ \solm.
By normalizing Eq. \eqref{enclosed_mass} to the entire enclosed mass $m_{enc}$, we find 
$m_{ext} /m_{enc}\lesssim 5\times 10^{-3} $ and 
$m_{com} /m_{enc} \lesssim 0.995$ while the sum of both terms equals unity.

\subsection{Could S4716 be a binary?}

Recent K-band observations with KECK suggest that the binary fraction in the Galactic
center and, in particular, the S-cluster is consistent with the local Solar neighborhood fraction \citep{Lefevre2009, Gautam2019}. Moreover, \citet{2016MNRAS.460.3494S} inferred a binary fraction of $\sim 70\%$ for the stellar population of the most recent star formation episode in the nuclear
star cluster $\sim 6$ Myr ago \citep[see e.g.][]{Lu2013}. Following the compelling evidence that stars form in multiple systems \citep[][]{Reipurth2000, Looney2000, Sadavoy2017}, we discuss the possibility that S4716 could be a binary system. However, as we will show, the parameter space of such a system is limited in order to be stable enough. Because the largest variability amplitude would be achieved for an eclipsing system, we will focus our discussion on these systems.
We note that the detectability of such a system in the S-cluster is at the limit of current instruments. This is underlined by the analysis of \cite{Gautam2019}, who report an uncertainty of $\rm \delta m_{K'}\,=\,0.03$ mag for the photometric study of the NSC. The authors found no evidence of an eclipsing binary in the S-cluster with a sample size of 17 stars. Due to the limited sample size, the existence of an eclipsing binary system cannot be ruled out considering an eclipsing binary fraction of approximately $\approx\,3\%$ \citep[][]{Pfuhl2014, Gautam2019}. 

For simplicity, we consider roughly equally luminous binary components. Then the variability amplitude in magnitudes is at most $\delta m\lesssim -2.5\log(1/2)\sim 0.75$ mag. Given the K-band magnitude uncertainty of $0.22$ mag this is just at the limit of detection.

For the pericenter distance of S4716 $r_{\rm p}\simeq 97.6\,{\rm AU}$, the component semi-major axis can be at most,
\begin{equation}
    a_{\rm bin}\lesssim 0.92 \left(\frac{r_{\rm p}}{97.6\,{\rm AU}} \right) \left(\frac{m_{\rm bin}}{3.20\,M_{\odot}}\right)^{1/3}\,{\rm AU}\,,
    \label{eq_abin}
\end{equation}
where the binary mass $m_{\rm bin}$ is set to the inferred mass of S4716. For the component distance greater than $a_{\rm bin}$, the binary would get disrupted at the periapse via the Hills mechanism. The corresponding orbital timescale of such a compact binary is
\begin{equation}
    P_{\rm bin}\lesssim
180.9\left(\frac{a_{\rm bin}}{0.92\,{\rm AU}}\right)^{3/2}\left(\frac{m_{\rm bin}}{3.20\,M_{\odot}}\right)^{-1/2}\,\text{days}\,.
    \label{eq_period_bin}
\end{equation}
If we consider the stellar diameter of $D_{\star}=2R_{\star}=2\times 2.45\,R_{\odot}=4.9\,R_{\odot}$, the eclipse event would last at most $t_{\rm ecl}\lesssim D_{\star}/v_{\rm bin}\sim 0.71 $ days, which is a dimming event too short to be practically detected with a sampling larger than one day. 

In other phases than the eclipse or for a non-eclipsing binary, components cannot be resolved out since the angular separation is $\lesssim 0.11$ mas, which is much smaller than the resolution limit of $\sim 60$ mas of eight-meter class telescopes. The detection by spectral disentangling would also be challenging since the maximum velocity shift of spectral lines would be $v_{\rm max}\sim v_{\rm bin}\simeq 55.5\,{\rm km\,s^{-1}}$ for an edge-on system. It could potentially be larger for more compact systems though, however, the timescales of the velocity shift would also shorten (for the separation of 0.1 AU, the orbital timescale would be just 6.5 days) 

In summary, S4716 could only be a very compact binary system with the separation of $\lesssim 1\,{\rm AU}$. The detection of such a compact system is extremely challenging in the crowded environment of the Galactic center since a potential eclipse event would last only $\sim 0.7$ days and the spectral line velocity shift would be at most $\sim 55.5\,{\rm km\,s^{-1}}$ for a given component separation.

\subsection{On the origin of the compact orbit}

The orbital compactness of S4716 with the semi-major axis of $a=1.94\pm 0.02\,{\rm mpc}=400\pm 4\,{\rm AU}$ is not straightforward to explain in terms of \textit{in situ} star-formation since the number density inside any cloud on these scales would have to be exceedingly large. This follows from the Roche density condition for the cloud to be tidally stable,
\begin{equation}
    n_{\rm Roche}\gtrsim 1.8\times 10^{15}\left(\frac{M_{\bullet}}{4\times 10^6\,M_{\odot}}\right) \left(\frac{r}{400\,{\rm AU}} \right)^{-3}\,{\rm cm^{-3}}\,.
\end{equation}
Therefore the occurrence of a star on such a compact orbit requires a certain migration mechanism that brings the star from larger distances closer in. One possibility is the inward migration within a dense accretion disc due to ``Type 1'' torques \citep{1980ApJ...241..425G}, which, however, works under the assumption that there has been a dense enough disc during the S-star lifetime.

A simpler explanation for the inward migration of an S star is the occurrence of other bodies in the S cluster that can effectively catalyze the semi-major axis and eccentricity changes. The Hills mechanism is the mechanism that involves a binary disruption. If the binary has the semi-major axis $a_{\rm bin}$ and the mass $m_{\rm bin}$, we obtain the tidal disruption radius that should coincide with the S4716 pericenter distance,
\begin{align}
    r_{\rm t}=r_{\rm p}&=a_{\rm bin}\left(\frac{M_{\bullet}}{m_{\rm bin}} \right)^{1/3} \notag\\
    &\sim 103\,\left(\frac{a_{\rm bin}}{1.4\,{\rm AU}} \right)\left(\frac{m_{\rm bin}}{10\,M_{\odot}} \right)^{-1/3}\,{\rm AU}\,.
\end{align}
In case S4716 would be the captured star from the binary disruption, its mean expected semi-major axis would be \citep{1991AJ....102..704H},
\begin{equation}
    <a_{\rm c}>\simeq 0.56a_{\rm bin} \left(\frac{M_{\bullet}}{m_{\rm bin}} \right)^{2/3}\sim 4256\,{\rm AU}\,.
\end{equation}
The corresponding eccentricity would be $e_{\rm c}=1-r_{\rm t}/<a_{\rm c}>\sim 0.976$. Hence, these orbital elements differ from those of S4716 -- the stellar orbit resulting from the binary disruption is expected to be much more elongated. Even for the initial binary mass of $m_{\rm bin}\sim 100\,M_{\odot}$ and the semi-major axis of $a_{\rm bin}\sim 2.9\,{\rm AU}$, the semi-major axis of the captured star would be $<a_{\rm c}>\simeq 1899\,{\rm AU}$ and its eccentricity of $e_{\rm c}=0.95$.

A more likely way to shorten the semi-major axis of an S star is the three-body interaction within the S cluster -- a close encounter of the star with a more massive perturber. If the original orbit is characterized by the orbital elements $a$, $e$, and $\iota$, the close encounter with a more massive star or the intermediate mass black hole (IMBH) will change the orbital elements to $a'$, $e'$, and $\iota'$, while the Jacobi intergral or the so-called Tisserand parameter is preserved \citep{2013degn.book.....M}. If the perturbing body has the semi-major axis $a_{\rm p}$, the Tisserand parameter can be calculated as,
\begin{equation}
    T(a,e,\iota)=\frac{a_{\rm p}}{a}+2\left[\frac{a}{a_{\rm p}}(1-e^2) \right]^{1/2}\cos{\iota}\,,
    \label{eq_tisserand}
\end{equation}
where $\iota$ is the inclination of an S star with respect to the perturber-SMBH plane. If the original orbit of S4716 was nearly circular, then the close encounter has to occur for $a\sim a_{\rm p}$ and a small inclination, hence $T\simeq 3$. Considering the current orbital elements of S4716, we can obtain $a_{\rm p}$ from the condition $T(a',e')\simeq 3$, where we assume that S4716 stays close to the orbital plane of the perturber. We obtain $a_{\rm p}\simeq 836\,{\rm AU}$ for $a'\simeq 400\,{\rm AU}$ and $e'=0.756$. With the apocenter distance of $r_{\rm a}=a'(1+e')\sim 702\,{\rm AU}$, S4716 would approach the mean distance of the perturber as expected. Hence, the encounter of S stars with more massive bodies such as IMBHs could shorten their semi-major axes by at least a factor of two and make their orbits more eccentric. The encounters could also be consecutive in an analogous way as comets interact with the large planets of the Solar system. This way S4716 could have migrated from the outer periphery of the S cluster all the way to its inner parts, though this needs to be verified in more detailed numerical simulations.

\section{Conclusion}
\label{sec:conclusion}

In this work, we present the observation of a new S-cluster member that we call S4716. This nomenclature follows the analysis presented \cite{Peissker2020d}. We furthermore reidentify S4711 in 2020 using the OSIRIS science cam. For S4716, we derive an orbital period of about 4 years. 
Inspecting the 2020 imaging data underlines the analysis of \cite{Peissker2020a}, \cite{Peissker2020d}, \cite{peissker2021b}, and this work, since we find within the uncertainties S62, S4711, S4716, and S29 at the expected position. We conclude that the periapse passage of $\rm \widetilde{S29}$ presented in GColl2022a is the result of the mixed observation of S62 and S4716.
In the following, we list our key findings and results from the discussion section.\newline
\begin{enumerate}
       \item S4716 can be observed with NACO, SINFONI, NIRC2, and OSIRIS between 2003 and 2020. The B8/9-V star orbits Sgr~A* in 4 years at distances between 98 AU and 702 AU. The K-band magnitude is about 17 mag and constant between 2003 and 2020 within the 1$\sigma$ uncertainty.
       \item S4716 is confused with S62 in 2004, 2011, and 2012. In 2016, the observation of S4716 is hampered due to S2. Furthermore, we find a new S-star called S148 that is traceable over multiple epochs. In 2017, we observe a blend star event for S148 with S4716. This event is underlined by the photometric analysis for both stars between 2016 and 2018.
       \item As a limit for the extended mass that may exist in addition to the compact mass of Sgr~A* within the S4716 orbit, we derive 2$\times$10$^{4}\,M_{\odot}$ (3$\sigma$).
       \item Calculated from the S4716 orbit, we find an enclosed mass of $(4.023\,\pm\,0.087)\,\times\,10^6\,M_{\odot}$ with a related distance of $8.028$ kpc $\pm\,2\%$ consistent with the literature \citep[][]{eht2022}.
       \item The data suggest that the claimed periapse passage of the S-star S29 in 2021 by GColl2022a is instead the periapse passage of the here presented newly discovered S-cluster member S4716.
\end{enumerate}

\acknowledgments

We are grateful for the comments by the anonymous referee that helped to improve this work. 
This work was supported in part by the
Deutsche Forschungsgemeinschaft (DFG) via the Cologne
Bonn Graduate School (BCGS), the Max Planck Society
through the International Max Planck Research School
(IMPRS) for Astronomy and Astrophysics as well as special
funds through the University of Cologne. Conditions and Impact of Star Formation is carried out within the Collaborative Research Centre 956, sub-project [A02], funded by the Deutsche Forschungsgemeinschaft (DFG) – project ID 184018867. AP, JC, SE, and GB contributed useful points to the discussion.
This research has made use of the Keck Observatory Archive (KOA), which is operated by the W. M. Keck Observatory and the NASA Exoplanet Science Institute (NExScI), under contract with the National Aeronautics and Space Administration. Michal Zaja\v{c}ek acknowledges the financial support by the GA\v{C}R EXPRO grant No. 21-13491X ``Exploring the Hot Universe and Understanding Cosmic Feedback''.

\bibliographystyle{aasjournal}
\bibliography{bib.bib}

\appendix

In this appendix, we will provide additional information about the analysis, follow up on the disputed identification of S29, and present supplementary data regarding the identification of S62 and S4716. We want to stress the observed blend star event in 2017.

\section{Used data}
\label{sec:data_appendix}
Here, we list the used data for this work (see Table \ref{tab:naco_data1} \& \ref{tab:data_sinfo3}). For the KECK observations, we download the data from the KOA archive. The analyzed KECK data was observed at 2006-05-03 (21 exposures, 58.8 sec total integration time) , 2019-04-20 (14 exposures, 39.2 sec total integration time), and 2019-06-30 (10 exposures, 28.0 sec total integration time). We want to highlight that this data was also analyzed in \cite{Parsa2017}, \cite{Peissker2019}, \cite{Peissker2020a}, \cite{Peissker2020b}, \cite{Peissker2020c}, \cite{Peissker2020d}, \cite{Ali2020}, \cite{peissker2021a}, \cite{peissker2021b}, and \cite{peissker2021c}.  
\begin{table*}[h!]
\centering
\begin{tabular}{ccccc}
\hline
\hline
\multicolumn{5}{c}{NACO}\\
\hline
Date (UT) & Observation ID & \multicolumn{1}{p{1.5cm}}{\centering number \\ of exposures }   & \multicolumn{1}{p{1.5cm}}{\centering Total \\ exposure time(s) } & \multicolumn{1}{p{1.5cm}}{\centering Seeing \\ in arcsec }  \\
\hline
2003-06-13 & 071.B-0078(A) & 102 & 510.00   & 0.6 \\
2005-07-25 & 075.B-0093(A) & 330 & 343.76  & 1.0  \\
2005-07-27 & 075.B-0093(C) & 158 & 291.09  & 0.9  \\
2005-07-29 & 075.B-0093(C) & 101 & 151.74  & 0.9  \\
2005-07-30 & 075.B-0093(C) & 187 & 254.07  & 0.9  \\
2005-07-30 & 075.B-0093(C) & 266 & 468.50  & 0.9  \\
2005-08-02 & 075.B-0093(C) & 80  & 155.77  & 0.9  \\
2007-03-04 & 078.B-0136(B) & 48  & 39.86   & 0.85  \\
2007-03-20 & 078.B-0136(B) & 96  & 76.19   & 0.85  \\
2007-04-04 & 179.B-0261(A) & 63  & 49.87   & 0.95  \\
2007-05-15 & 079.B-0084(A) & 116 & 181.88  & 0.75  \\
2008-02-23 & 179.B-0261(L) & 72  & 86.11  & 0.9   \\
2008-03-13 & 179.B-0261(L) & 96  & 71.49  & 0.8   \\
2008-04-08 & 179.B-0261(M) & 96  & 71.98  & 0.8   \\
2010-03-29 & 183.B-0100(L) & 96  & 74.13   & 0.9  \\
2010-05-09 & 183.B-0100(T) & 12  & 16.63   & 0.85  \\
2010-05-09 & 183.B-0100(T) & 24  & 42.13   & 0.85  \\
2010-06-12 & 183.B-0100(T) & 24  & 47.45   & 0.85  \\
2010-06-16 & 183.B-0100(S) & 48  & 97.78   & 0.65  \\
2013-09-01 & 091.B-0183(B) & 178 & 1780.00 & 1.05  \\
2016-03-22 & 594.B-0498(I) & 144 & 6300    & 0.6  \\
2016-09-25 & 097.B-0216(B) & 83 &  16.60   & 0.75  \\
2016-09-26 & 594.B-0498(K) & 13 &  52.00   & 0.7  \\
2017-06-16 & 598.B-0043(L) & 36  & 144.00  & 0.65  \\
2018-04-24 & 101.B-0052(B) & 120 & 1200.00 & 0.55  \\
2019-09-27 & 5102.B-0086(H) & 41 & 164.00  & 0.65  \\

\hline  
\end{tabular}
\caption{NACO K-band data observed between 2004-2019. Because several individual observations are stacked to a final mosaic, we list a range of the seeing for the corresponding used on-source exposures.Here, seeing is refering to the ESO archive keyword {\it DIMM-Seeing-at-start}. We note that observations with a seeing greater than 1.5 arcsec are excluded because the identification of individual sources is hampered.}
\label{tab:naco_data1}
\end{table*}

\begin{table*}[htbp!]
        \centering
        \begin{tabular}{ccccccc}
        \hline\hline
        \\      Date & Observation ID  & \multicolumn{3}{c}{Amount of on source exposures} & Exp. Time & Seeing \\  \cline{3-5} &  & Total & Medium & High &  & \\
        (YYYY:MM:DD) &  &  &  &  & (s) & (arcsec) \\ \hline\hline 
        
        2013-04-06 & 091.B-0088(A)  & 8 &   0  &  8  &     600 & 0.7  \\
        2013-04-07 & 091.B-0088(A)  & 3 &   0  &  3  &     600 & 0.6  \\
        2013-04-08 & 091.B-0088(A)  & 9 &   0  &  6  &     600 & 1.3  \\
        2013-04-09 & 091.B-0088(A)  & 8 &   1  &  7  &     600 & 0.7  \\
        2013-04-10 & 091.B-0088(A)  & 3 &   0  &  3  &     600 & 0.9  \\
        2013-08-28 & 091.B-0088(B)  & 10 &  1  &  6  &     600 & 0.7 \\
        2013-08-29 & 091.B-0088(B)  & 7 &  2   &  4  &     600 & 0.9 \\
        2013-08-30 & 091.B-0088(B)  & 4 &  2   &  0  &     600 & 0.9 \\
        2013-08-31 & 091.B-0088(B)  & 6 &  0   &  4  &     600 & 0.7 \\
        2013-09-23 & 091.B-0086(A)  & 6 &   0  &  0  &     600 & 1.3  \\
        2013-09-25 & 091.B-0086(A)  & 2 &   1  &  0  &     600 & 1.1  \\
        2013-09-26 & 091.B-0086(A)  & 3 &   1  &  1  &     600 & 0.7  \\   
        2014-02-27 & 092.B-0920(A) &  4 &   1  &  3  &     600 & 0.8   \\
        2014-02-28 & 091.B-0183(H) &  7 &   3  &  1  &     400 & 0.9   \\
        2014-03-01 & 091.B-0183(H) & 11 &   2  &  4  &     400 & 0.8  \\
        2014-03-02 & 091.B-0183(H) &  3 &   0  &  0  &     400 & 0.9   \\
        2014-03-11 & 092.B-0920(A) & 11 &   2  &  9  &     400 & 1.0   \\
        2014-03-12 & 092.B-0920(A) & 13 &   8  &  5  &     400 & 0.9   \\
        2014-03-26 & 092.B-0009(C) & 9  &   3  &  5  &     400 & 0.9   \\
        2014-03-27 & 092.B-0009(C) & 18 &   7  &  5  &     400 & 1.0   \\
        2014-04-02 & 093.B-0932(A) & 18 &   6  &  1  &     400 & 0.9    \\
        2014-04-03 & 093.B-0932(A) & 18 &   1  &  17 &     400 & 0.8    \\
        2014-04-04 & 093.B-0932(B) & 21 &   1  &  20 &     400 & 0.8   \\
        2014-04-06 & 093.B-0092(A) &  5 &   2  &  3  &     400 & 0.7  \\
        2014-04-08 & 093.B-0218(A) & 5  &   1  &  0  &     600 & 1.1   \\
        2014-04-09 & 093.B-0218(A) &  6 &   0  &  6  &     600 & 0.8   \\
        2014-04-10 & 093.B-0218(A) & 14 &   4  &  10  &    600 & 0.8   \\
        2014-05-08 & 093.B-0217(F) & 14 &   0  &  14  &    600 & 0.7   \\
        2014-05-09 & 093.B-0218(D) & 18 &   3  &  13  &    600 & 0.8  \\
        2014-06-09 & 093.B-0092(E) &  14 &   3  &  0  &    400 & 0.8   \\
        2014-06-10 & 092.B-0398(A)/093.B-0092(E) & 5 &   4  &  0   & 400/600 & 0.7 \\
        2014-07-08 & 092.B-0398(A)  & 6 &   1  &  3   &    600 & 0.7 \\
        2014-07-13 & 092.B-0398(A)  & 4 &   0  &  2   &    600 & nan \\
        2014-07-18 & 092.B-0398(A)/093.B-0218(D)  & 1 &   0  &  0   &    600 & 1.1 \\
        2014-08-18 & 093.B-0218(D)  & 2 &   0  &  1   &    600  & 1.1 \\ 
        2014-08-26 & 093.B-0092(G)  &  4 &   3  &   0 &    400  & 1.3  \\
        2014-08-31 & 093.B-0218(B)  & 6 &   3   &   1 &    600  & 0.9 \\
        2014-09-07 & 093.B-0092(F)  & 2 &   0  &  0  &     400  & 0.8  \\
        2015-04-12 & 095.B-0036(A)  & 18 &  2 & 0 &        400  & 0.9 \\
        2015-04-13 & 095.B-0036(A)  & 13 &  7 & 0 &        400  & 0.9 \\
        2015-04-14 & 095.B-0036(A)  & 5  &  1 & 0 &        400  & 0.9 \\
        2015-04-15 & 095.B-0036(A)  & 23 &  13  & 10 &     400  & 0.9 \\
        2015-08-01 & 095.B-0036(C)  & 23 &   7  & 8  &     400  & 1.3 \\
        2015-09-05 & 095.B-0036(D)  & 17 &  11  & 4  &     400  & 1.5 \\

        \hline  \\
        \end{tabular}
        
        \caption{SINFONI data observed in 2013, 2014, and 2015. From the H+K data cubes, we extract the continuum images covering the wavelength range of $2.0-2.2\,\mu m$. We furthermore indicate the related seeing for every observation night. Because seeing changes for every single observation, we provide an averaged value that represents the overall quality of the observation night. We note that also under bad conditions (seeing larger than 1.5 arcsec), high-quality observations are, in principle, possible.}
        \label{tab:data_sinfo3}
        \end{table*}

In Table \ref{tab:positions_instruments_band}, we indicate the instruments related to the data points shown in Fig. \ref{fig:orbit} and Fig. \ref{fig:on_sky}. Although the filters used in this analysis cover slightly different spectral ranges, the spectroscopic offset is rather small in terms of absolute numbers. Because we assume that the magnitude ratio between stars is constant for the investigated sample, the absolute sensitivity should not be a major contributing factor to the uncertainties. Since we use S2 as a reference star with a related magnitude of 14.1 mag in the K-band, the ratio of another star should be independent of the used instrument if the frame (sky correction, on-source integration time, detector health) of the observations is comparable. Consult Sec. \ref{sec:results} for an additional discussion.

\begin{table*}[htb]
\centering
\begin{tabular}{cccccccc}\hline \hline
Epoch & $\Delta$R.A. (as) & $\Delta$DEC. (as)  & $\Delta$R.A. error (as) &$\Delta$DEC. error (as) & $\lambda$ [$\mu m$] & Resolution (as) & Instrument\\ \hline 
 
2003.44 & -0.041  &     0.066 &  0.007 & 0.007    & 1.97-2.32 &   0.039   & NACO \\
2005.26 & -0.033  &    -0.002 &  0.007 & 0.007    & 1.97-2.32 &   0.046   & NACO \\
2005.58 &  0.019  &     0.033 &  0.007 & 0.007    & 1.97-2.32 &   0.039   & NACO \\
2006.33 & -0.015  &     0.077 &  0.007 & 0.007    & 1.94-2.29 &   0.029   & NIRC2 \\
2007.25 & -0.039  &     0.073 &  0.007 & 0.007    & 1.97-2.32 &   0.030   & NACO \\
2008.46 & -0.053  &     0.045 &  0.007 & 0.007    & 1.97-2.32 &   0.033   & NACO \\
2010.24 & -0.017  &     0.079 &  0.013 & 0.013    & 1.97-2.32 &   0.039   & NACO \\
2013.40 &   0.006 &   -0.013  &  0.007 & 0.007    & 2.00-2.20 &   0.037   & SINFONI \\
2013.66 &  0.013  &     0.000 &  0.007 & 0.007    & 1.97-2.32 &   0.046   & NACO \\
2014.50 &  -0.024 &     0.070 &  0.007 & 0.007    & 2.00-2.20 &   0.037   & SINFONI \\
2015.30 &  -0.050 &     0.062 &  0.007 & 0.007    & 2.00-2.20 &   0.042   & SINFONI \\    
2019.30 & -0.027  &     0.073 &  0.007 & 0.007    & 1.97-2.32 &   0.029   & NIRC2\\
2019.67 & -0.044  &     0.064 &  0.005 & 0.005    & 1.94-2.29 &   0.028   & NIRC2\\
2019.73 & -0.053  &     0.066 &  0.007 & 0.007    & 1.97-2.32 &   0.038   & NACO\\
2020.42 & -0.050  &     0.050 &  0.005 & 0.005    & 2.12-2.22 &   0.039   & OSIRIS \\
2020.58 & -0.050  &     0.050 &  0.005 & 0.005    & 2.12-2.22 &   0.029   & OSIRIS \\
2021.24 & -0.051  &     0.034 & 0.0005 & 0.0005   & 2.00-2.40 &   $65\times 10^{-6}\,^{\star}$   & GRAVITY\\
2021.41 & 0.0029  &  -0.01176 & 0.0005 & 0.0005   & 2.00-2.40 &   $65\times 10^{-6}\,^{\star}$   & GRAVITY\\
2021.47 & 0.0129  &   -0.0029 & 0.0005 & 0.0005   & 2.00-2.40 &   $65\times 10^{-6}\,^{\star}$   & GRAVITY\\
2021.56 & 0.0182  &   0.00941 & 0.0005 & 0.0005   & 2.00-2.40 &   $65\times 10^{-6}\,^{\star}$   & GRAVITY\\
\hline \hline
\end{tabular}
\caption{List of derived positions and the corresponding uncertainty for S4716. Using a GAUSSIAN fit to derive the position of S4716 is accompanied by an underestimated uncertainty that does not reflect the noise character of the region \citep[][]{peissker2021c}. An uncertainty of $\pm\,5-7\times 10^{-3}$as reflects positional uncertainties in combination with crowding challenges in the S-cluster. In addition, we indicate the resolution of the related data by measuring the FWHM of S2. For GRAVITY data, we use the astrometric accuracy indicated by GColl2022a.}
\label{tab:positions_instruments_band}
\end{table*}

\section{MCMC statistics}
\label{sec:mcmc}

\begin{figure*}[htbp!]
	\centering
	\includegraphics[width=1.\textwidth]{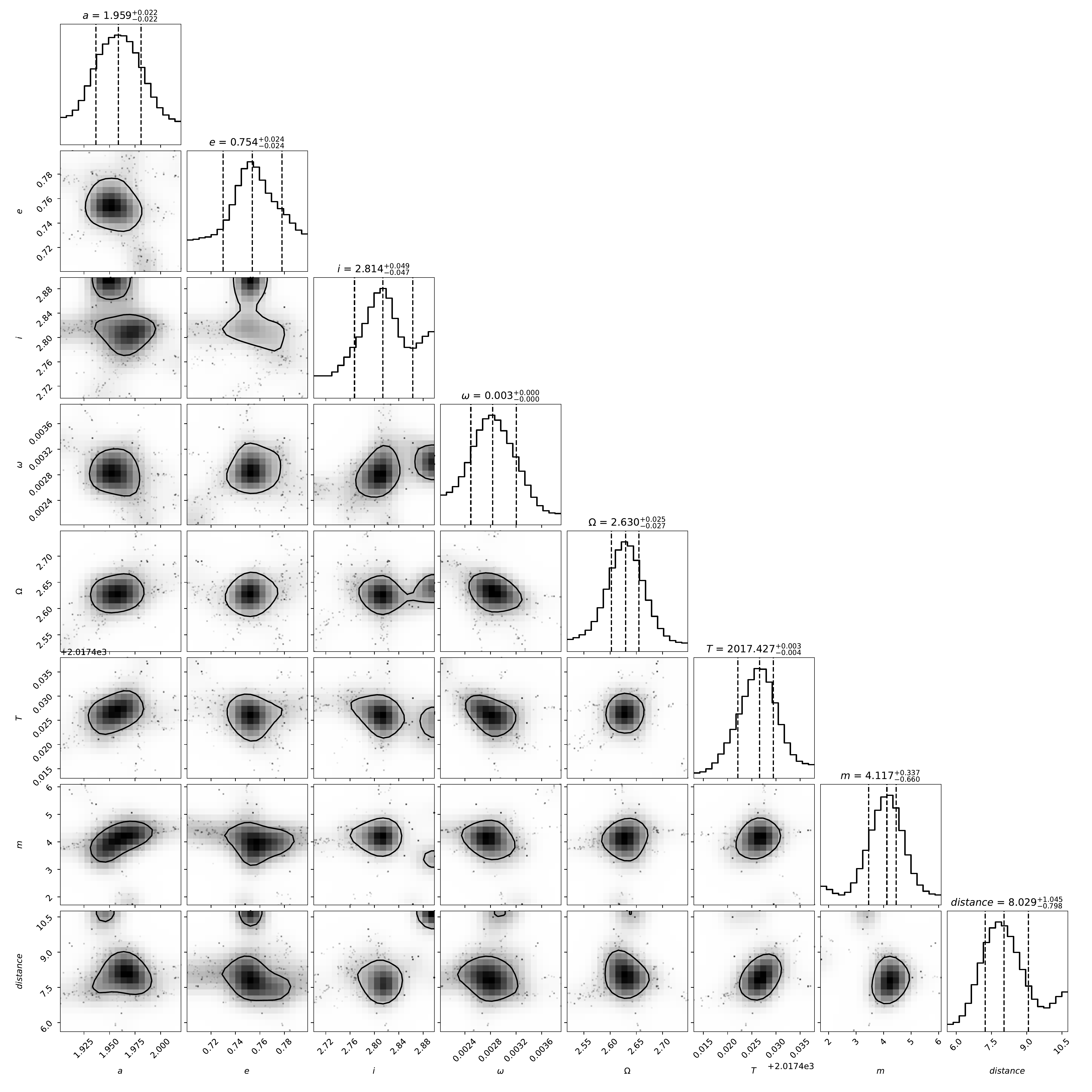}
	\caption{MCMC simulation of the orbital elements given in Table \ref{tab:orbit_elements} for S4716. Overall, we find a convincing agreement with the input parameters which is reflected in the compact shape of the shown output.}
\label{fig:mcmc}
\end{figure*}

Here, we present the outcome of the MCMC simulations (Fig. \ref{fig:mcmc}). A compact distribution indicates a well-defined input parameter.
For transparency, we are aware of the second peak for the inclination close to $\rm i=2.90=166.1^{\circ}$. Of course, this implies that other solutions to the Keplerian fit simulations may also produce satisfying results. Nevertheless, we would like to guide the reader to Fig. \ref{fig:on_sky} where we show the Keplerian solution for the observed data points for S4716. Since the Keplerian solution reasonably matches the observations, we use the result for the inclination from the MCMC statistics to determine a proper uncertainty range. For the uncertainty, we use the mean error with $\rm i_{mean\,err}\,=\,\frac{166.1^{\circ}-161.2^{\circ}}{2}\,=\,2.4^{\circ}$. With a mean value for the inclination of $\rm i_{mean}\,=\,163.6^{\circ}$, we estimated a final value of $\rm 163.6^{\circ}\,\pm\,2.4^{\circ}$.    

\section{Mass and distance of Sgr~A*}
\label{sec:mass_distance_mcmc}

\begin{figure}[htbp!]
	\centering
	\includegraphics[width=.5\textwidth]{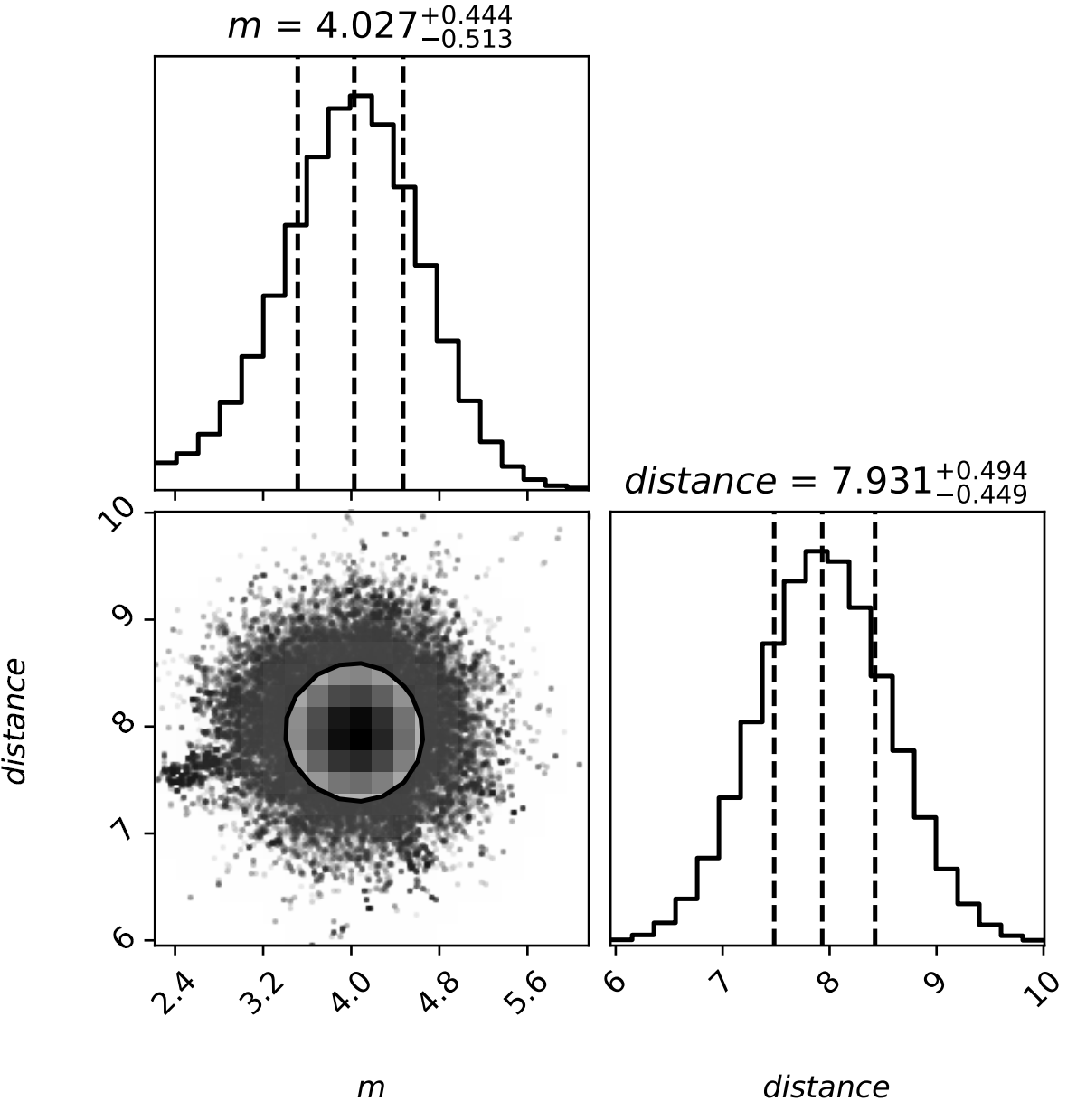}
	\caption{Results of the first MCMC run. As described in the text, we use $10^4$ iteration steps. Because of the open boundaries, the range of possible solutions for the mass and the distance covers almost 10$\%$ of the peak value.}
\label{fig:mass_distance_1}
\end{figure}

\begin{figure}[htbp!]
	\centering
	\includegraphics[width=.5\textwidth]{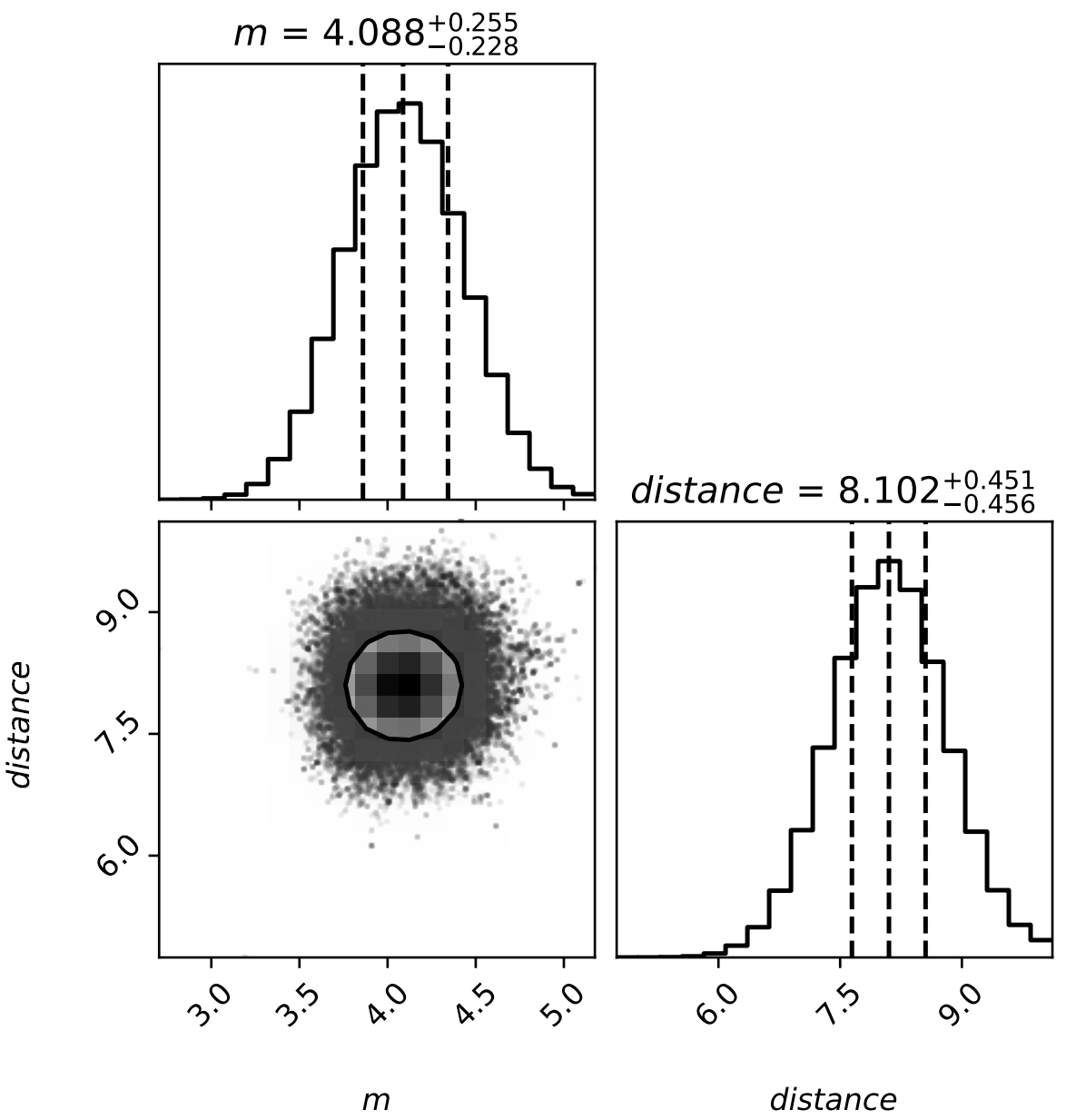}
	\caption{Results of the second MCMC run. Like the results presented in Fig. \ref{fig:mass_distance_1}, we use $10^4$ iterations steps for the MCMC simulations. However, here the uncertainty range is decreased to $5\%$.}
\label{fig:mass_distance_2}
\end{figure}

\begin{figure}[htbp!]
	\centering
	\includegraphics[width=.5\textwidth]{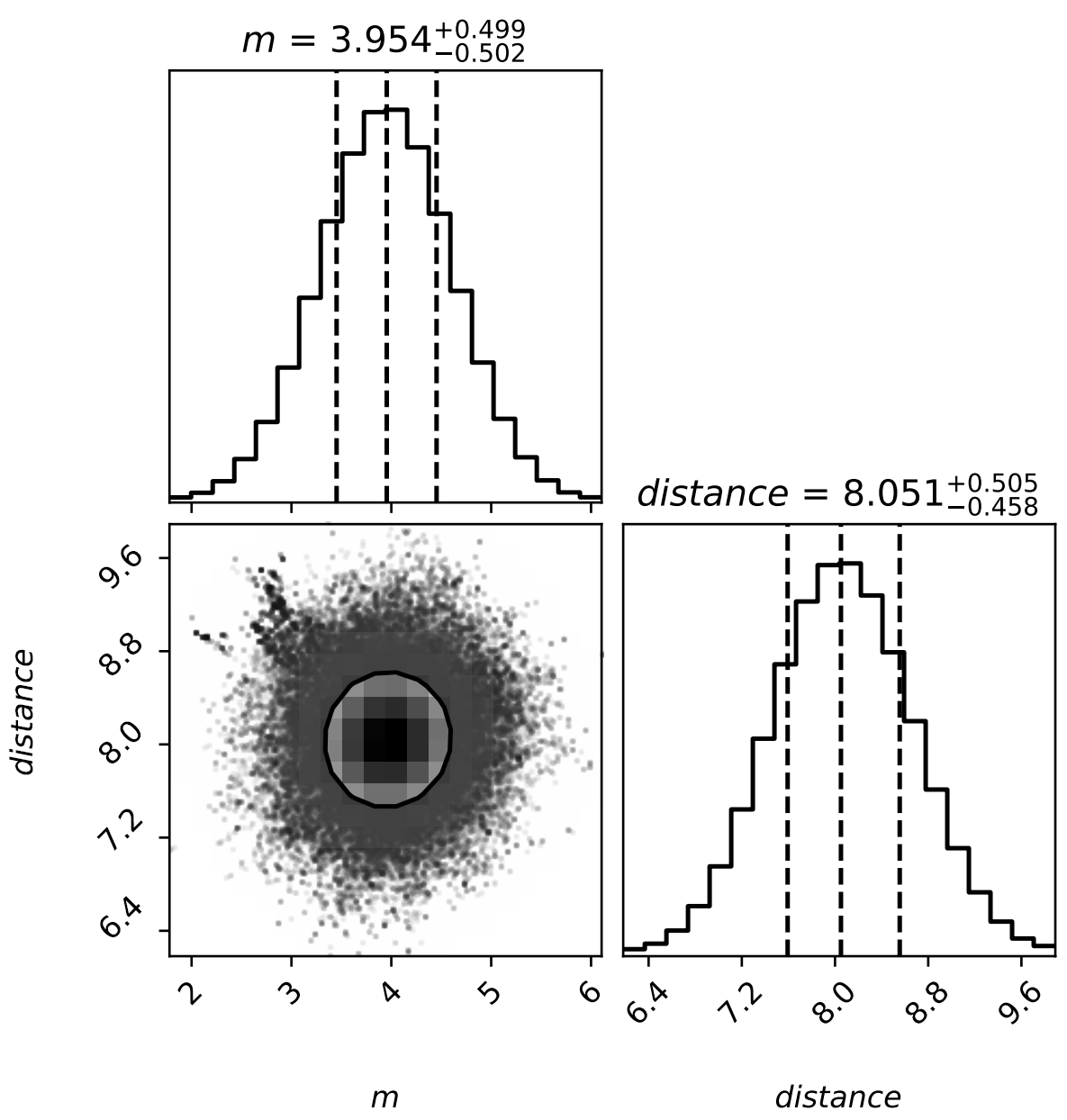}
	\caption{Results of the third MCMC run. Here, the uncertainty range is about $10\%$ like the results presented in Fig. \ref{fig:mass_distance_1}.}
\label{fig:mass_distance_3}
\end{figure}

Since we executed the MCMC simulations with iteration steps of i=10000 several times, we noticed a strong correlation of the mass and distance for Sgr~A*. The results of these runs do show agreeing values indicating that S4716 is a perfect candidate for the evaluation of the enclosed mass and the distance of the SMBH. In the following, we present the related figures of the MCMC runs (Fig. \ref{fig:mass_distance_1}-\ref{fig:mass_distance_3}).

\section{Line of sight velocity of S4716}
\label{sec:spec_los}
\begin{figure}[htbp!]
	\centering
	\includegraphics[width=.5\textwidth]{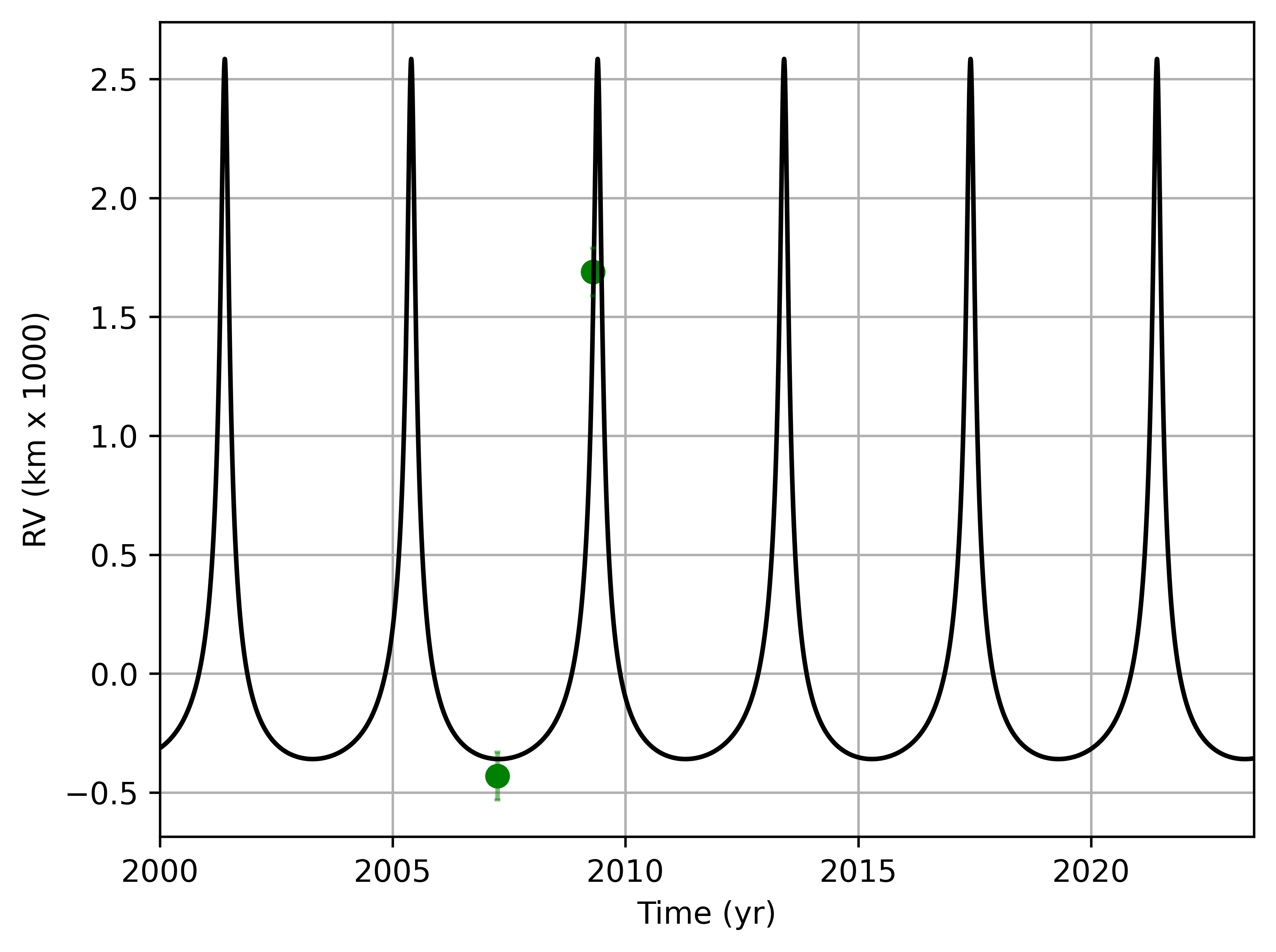}
	\caption{Line of sight velocity for S4716. Here we used SINFONI data of 2007 and 2009 to derive the Doppler-shifted LOS velocity of S4716 with the Br$\gamma$ absorption line. The green data points are not fitted and represent an independent parameter to confirm the here presented analysis.}
\label{fig:los}
\end{figure}

\begin{figure}[htbp!]
	\centering
	\includegraphics[width=.5\textwidth]{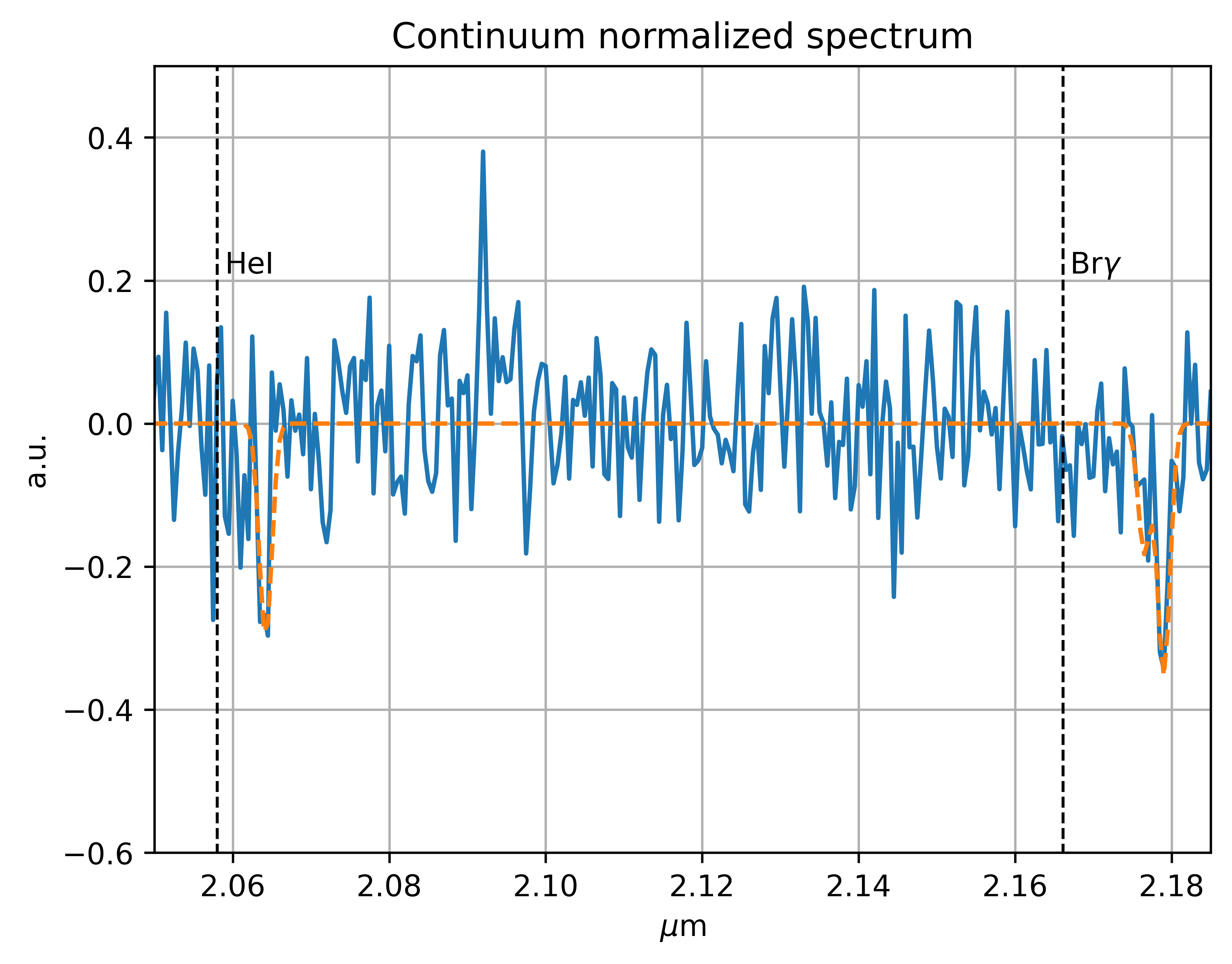}
	\caption{Spectrum of S4716 in 2009 as observed with SINFONI. In contrast to 2007 (Fig. \ref{fig:spec}), the velocity is redshifted with about 1700 km/s with respect to the Br$\gamma$ restwavelength at 2.1661$\mu m$.}
\label{fig:spec_4716_2009}
\end{figure}

Here we show the related LOS velocity to the spectrum shown in Fig. \ref{fig:spec}. The fit presented in Fig. \ref{fig:los} is purely based on the outcome of the Keplerian orbit presented in Fig. \ref{fig:orbit}. This means, we derive the orbital solution and the related LOS evolution along with the orbit independently from the estimated Dopplershifted velocity of almost -500 km/s in 2007 and 1690 km/s in 2009 (Fig. \ref{fig:spec_4716_2009}). 
With this approach, we have an independent parameter to verify the orbital solution and the related source detection for these epochs. 

\section{Comment on the detection of S62 in 2019}
\label{sec:comment_s29}
While we highly appreciate the attempt of observing S62 with GRAVITY in 2019 GColl2021a, we would like to address some challenges regarding the interpretation of the presented data. In Tab. \ref{tab:positions_S29_S62_2019}, we present the projected positions of S29$_{Gillessen}$ (G17), S62 \citep[][]{Peissker2020a}, and the GRAVITY star that we denote with $\widetilde{S29}$ observed in 2019 and analyzed in GColl2021a.
\begin{table*}[hp!]
\centering
\begin{tabular}{ccccccc}
\hline
\hline
\multicolumn{5}{c}{Projected positions of S29$\rm _{Gillessen}$, S62, and $\widetilde{S29}$ in 2019}\\
\hline
Epoch & S29$\rm _{Gillessen\,\,x}$ & S29$\rm _{Gillessen\,\,y}$ & $S62_x$ & $S62_y$ & $\widetilde{S29}_x$ & $\widetilde{S29}_y$\\
\hline
2019.66 & -98.6 & +108.6 & -85.3 & +97.2 & -87.6 & +92.5 \\ 
2019.75 & -97.3 & +105.3 & -84.4 & +96.7 & -85.9 & +88.7\\ 
\hline  
\end{tabular}
\caption{Positions of S29$\rm _{Gillessen}$, S62, and $\widetilde{S29}$ observed in 2019. We extract these values from the related publications (GColl2021a, GColl2022a) and show a visualization of this table in Fig. \ref{fig:s62_s29_mock_image}. We would like to note that there is a striking distinction between the positions of S29$\rm _{Gillessen}$ and $\widetilde{S29}$.}
\label{tab:positions_S29_S62_2019}
\end{table*}
Considering the close distance of S62 to $\widetilde{S29}$ in 2019 results in a challenging interpretation, to connect the latter star to S29$\rm _{Gillessen}$. To visualize this point, we created a mock image with the FOV of GRAVITY and the positions of the three stars in 2019.66 and 2019.75 as listed in Table \ref{tab:positions_S29_S62_2019}.
\begin{figure}[htbp!]
	\centering
	\includegraphics[width=1.\textwidth]{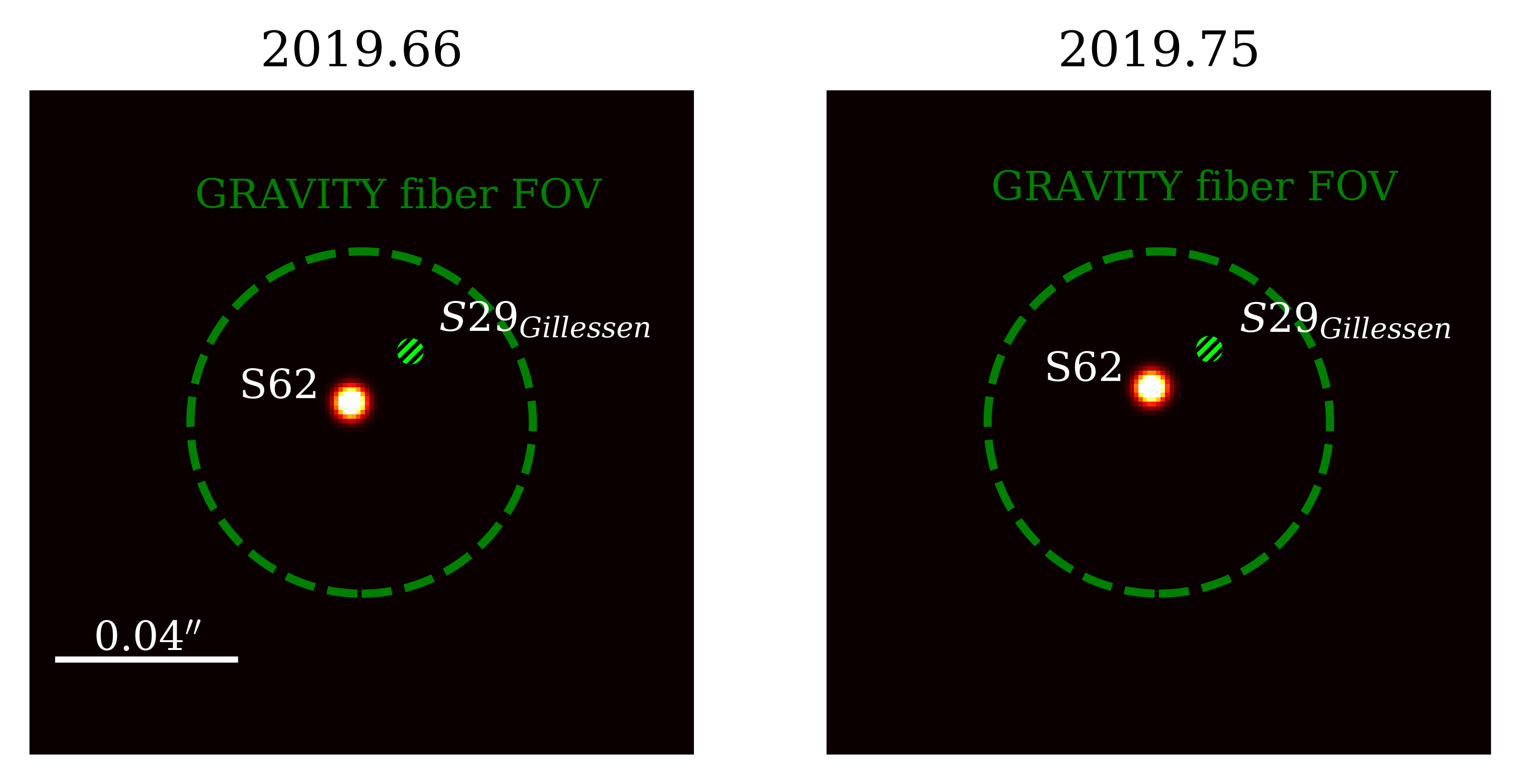}
	\caption{Visualization of the source confusion as summarized in Table \ref{tab:positions_S29_S62_2019} in 2019.66 and 2019.75. Here, the green dashed circle indicates the fiber FOV of GRAVITY where the authors of \cite{GravityCollaboration2021} claim the observation of a stellar source that is located in the center of these images. As it is implied by this figure, the closest source to the center of the GRAVITY fiber FOV is S62. The predicted position of S29$_{\rm Gillessen}$ based on compromised orbital elements taken from G17 \citep[see clarifying comments in][]{Peissker2020a} is at too large a distance from the center of the GRAVITY fiber FOV in order to still be considered. It is important to note that no bright source was located at the predicted S29$_{\rm Gillessen}$ in 2019.66 and 2019.75. Hence, we use a lime and black cross hatched symbol. Consequently we are allowed to raise the question, why the authors of \cite{GravityCollaboration2021} came to the conclusion of neglecting the confusion free identification of S62. This misidentification is a key element that gave rise to the S29 orbit as published now in \cite{GravityCollaboration2022a}. We would like to express our doubts about this identification since the authors pointed GRAVITY to the position of S62 based on our proposal \citep[see][]{peissker2019ATel} and private communication to the GRAVITY observing team on the Paranal mountain. Here, the figure size is 150$\times$150 mas, the diameter of the dashed green circle is about 74 mas.}
\label{fig:s62_s29_mock_image}
\end{figure}
The authors of GColl2021a argue that the proper motion of the observed GRAVITY star does not match S62 but rather S29$\rm _{Gillessen}$. The authors ignore the predicted periapse passage for S29$\rm _{Gillessen}$ but more importantly, the increasing distance to the observed GRAVITY star. Compared to S62, S29$\rm _{Gillessen}$ is over $400\%$ away from the GRAVITY star. Therefore, we propose that S62 = $\widetilde{S29}$ =  GRAVITY star in 2019. 

\section{The detection of S62}

\begin{figure}[htbp!]
	\centering
	\includegraphics[width=1.\textwidth]{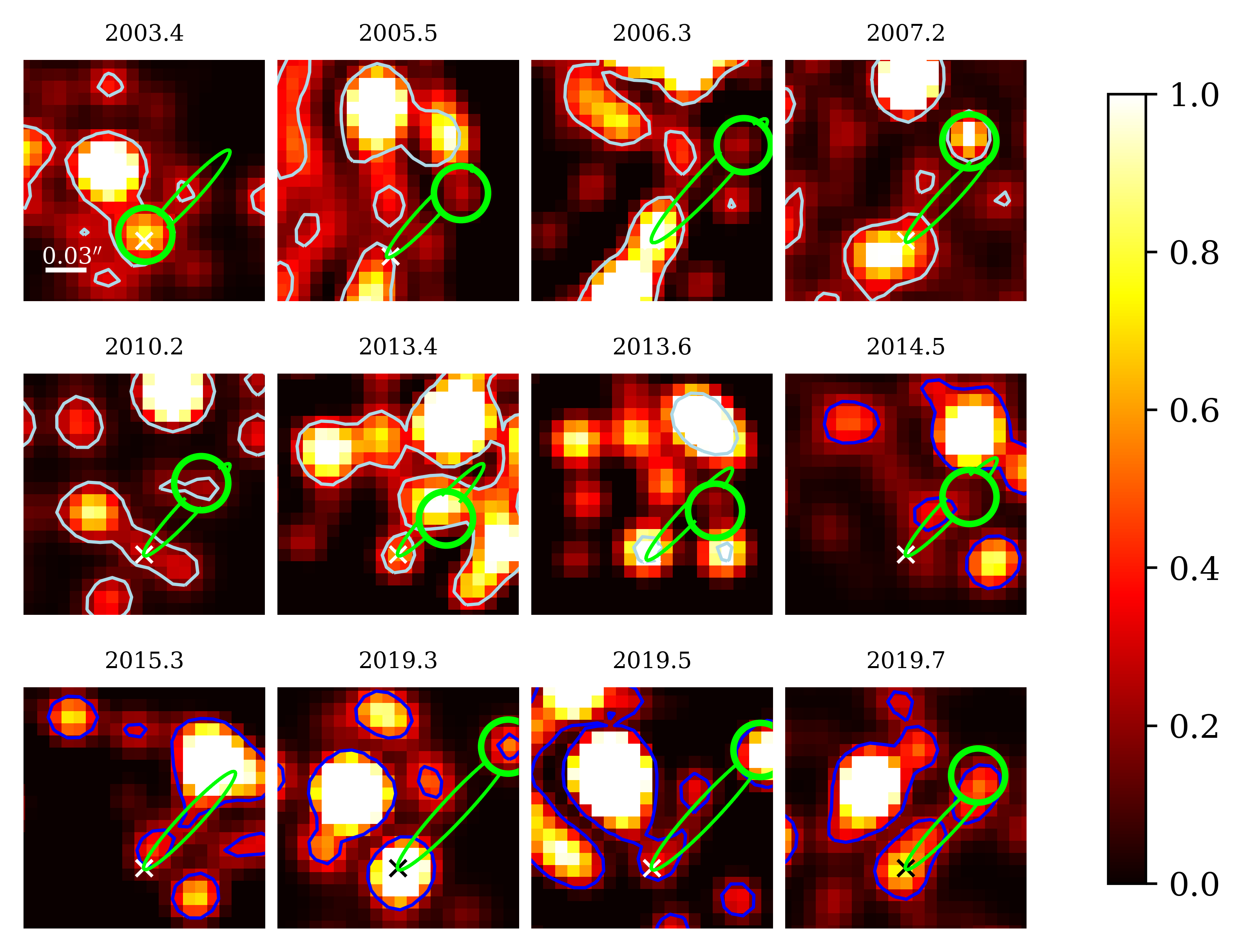}
	\caption{Detection of S62 in the here presented data set. The observations are carried out with the VLT (NACO and SINFONI) and the KECK (NIRC2) telescope. In 2015, S62 is confused with S2. Likewise in Fig. \ref{fig:timeline}, the data observed with NIRC2/KECK for the epochs 2006.3, 2019.3, and 2019.5. We furthmore use SINFONI/VLTA data for the epochs 2013.4, 2014.5, and 2015.3. The remaining epochs are observed with NACO/VLT. Magnitude variations do not the represent the intrisic flux of the star. These variations are related to the background, data quality, nearby sources, and the contrast (which is choosen to show the surrounding single stars and S62). A detailed analysis of the K-band magnitude of S62 can be found in \cite{peissker2021b}.}
\label{fig:s62_timeline}
\end{figure}

In this section, we show the same data as presented in Fig. \ref{fig:timeline} but with the S62 orbit. Even though we have presented numerous observations of this star \citep[][]{Peissker2020a, Peissker2020d, peissker2021c}, we revisited the available data with the approach of using a combined PSF (APSF+airy ring model) for the Lucy Richardson deconvolution algorithm. As expected, the data shown in Fig. \ref{fig:s62_timeline} clearly shows the star S62 at the expected position on its orbit around Sgr~A*.

\section{Observation of S4716 in 2016, 2017, and 2018}
\label{sec:s4716_2016_2017_2018}
\begin{figure}[htbp!]
	\centering
	\includegraphics[width=1.\textwidth]{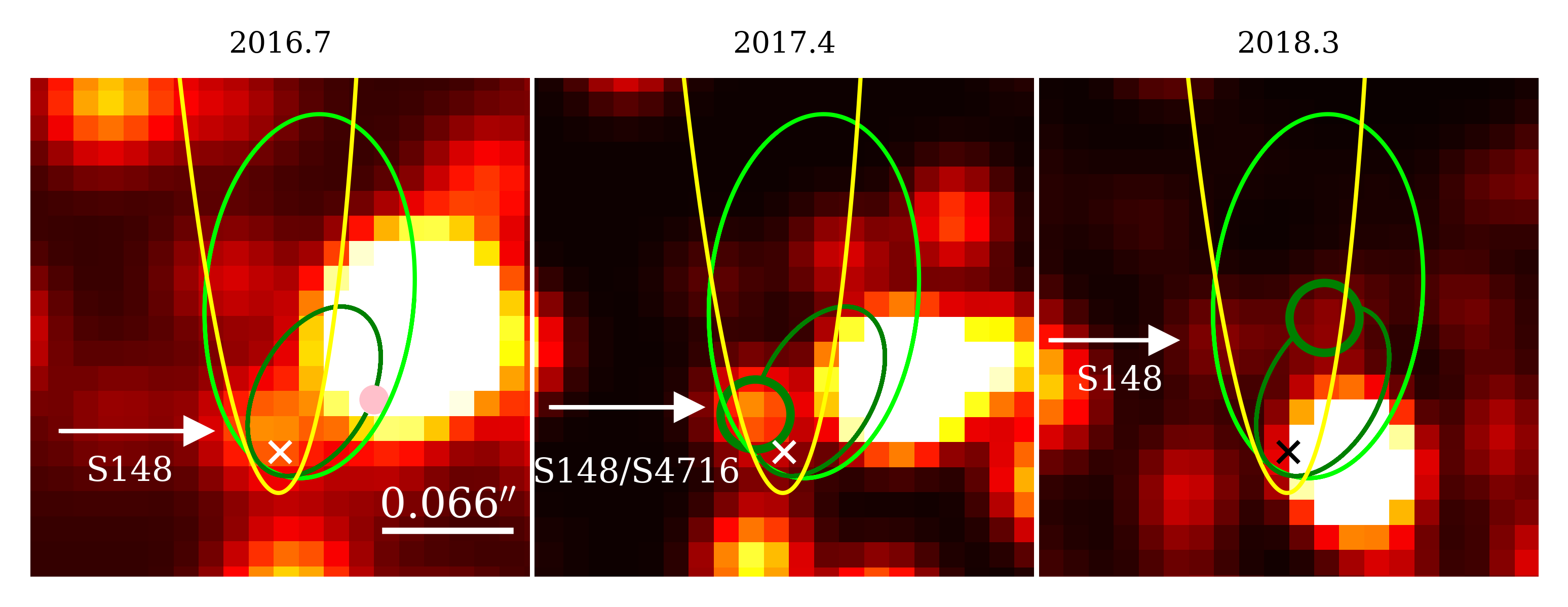}
	\caption{K-band view on the environment of Sgr~A* (indicated by a $\times$) observed with NACO showing S2 and S4716 for 2016.7, 2017.4, and 2018.3. In 2016, we illustrate the expected position of S4716 in the related epoch with a pink colored patch on the green orbit. The S2 orbit is shown in lime. The yellow trajectory shows the orbit of S148. North is up, east is to the left.}
\label{fig:interference_s4716}
\end{figure}

With a K-band magnitude of 14.15 mag, the bright S-cluster member S2 is the dominant source in the cluster \citep[][]{Schoedel2002, Habibi2017}. We already faced the observational influence of S2 regarding S62 in 2015 \citep[][]{Peissker2020a}. For the same epoch, the projected position of S2 comes close to the apoapse of S4716 (Fig. \ref{fig:timeline} and Fig. \ref{fig:s62_timeline}). Therefore, we will investigate the gap between 2016 and 2018 for the possible detection of S4716 besides the influence of S2 (see Fig. \ref{fig:interference_s4716}). We would like to note that the S-cluster is constantly monitored. Due to blending, confusion, variable background, and general data quality related challenges, the number of reported stars in the literature (see, e.g., G17) is much lower than the observed stars \citep[see][]{Peissker2020d}. Therefore, we will introduce a new S-star that crosses the orbit of S4716. We call this new identified star S148, following the widely accepted nomenclature \citep[][]{Ali2020}. In Sec. \ref{sec:s148}, we will elaborate on the analysis of S148.
We find that S4716 is confused with S148 in 2017 (see the circled star in Fig. \ref{fig:interference_s4716} in 2017.4). The K-band magnitude of S148 is almost 1 mag higher in 2017 (16.0 mag) compared to 2016 (16.9 mag), underlining the confusion with S4716. In contrast, the circled star in 2018.3 (Fig. \ref{fig:interference_s4716}) shows a magnitude of 17.0 mag, indicating the reidentification of S4716 (see Sec. \ref{sec:parameters_s4716}) in this epoch. This followed by the presented identification in 2019 (Fig. \ref{fig:timeline}) and 2020 (Fig. \ref{fig:ident_2020}). 

\section{The newly identified S-cluster star S148}
\label{sec:s148}
\begin{figure}[htbp!]
	\centering
	\includegraphics[width=.8\textwidth]{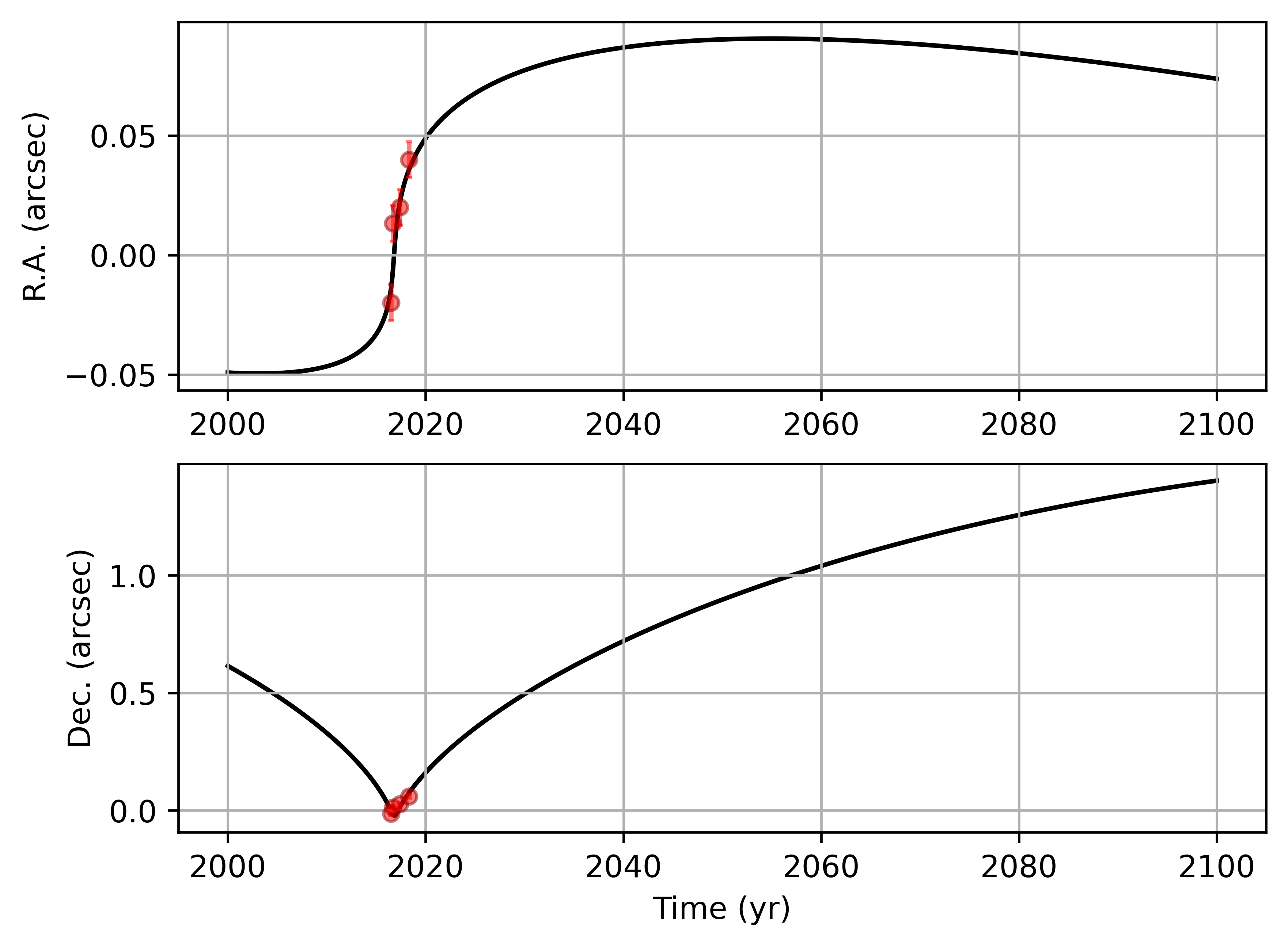}
	\caption{Orbital solution for S148 based on the measurements in 2016.5, 2016.7, 2017.4, and 2018.3. Because of the increasing distance between S148 and S2, this solution is likely to be improved in the near future.}
\label{fig:s148_orbit}
\end{figure}

Here, we will present the orbital elements of S148. In 2016, S148 passed Sgr~A* (see Fig. \ref{fig:interference_s4716}) and is located close (about 12.5-25 mas) to S63. In agreement with the presented list for verifying newly identified S-stars (Sec. \ref{appendix:checklist}) we find this star in 3 consecutive years on a Keplerian orbit at the expected position in 2016.5, 2016.7, 2017.4, and 2018.3 using the LR algorithm. In 2018.3, we additionally observe S148 using a SM-filter (see Sec. \ref{fig:filter}).\newline  
The orbital elements are displayed in Tab \ref{tab:orbit_elements_s148}, the solution for the Kepler fit is shown in Fig. \ref{fig:s148_orbit}. The uncertainties are adapted to account for the limited orbit coverage of the data because of confusion with S2 before 2016 and after 2018.
\begin{table*}[htb]
\centering
\begin{tabular}{ccccccc}\hline \hline
Source  & $a$ (mpc)&	$e$  &	$i$($^o$)&$\omega$($^o$)&	$\Omega$($^o$)&	$t_{\rm closest}$(yr)\\ \hline 
S148    &31.13 $\pm$ 0.1   &0.971 $\pm$  0.05 &111.15 $\pm$ ~5.00  &~174.75 $\pm$  5.00  &359.24 $\pm$~5.00  &2016.8 $\pm$ 0.2\\
\hline \hline
\end{tabular}
\caption{Orbital elements of S148. The uncertainties are an upper limit and reflect the poor data baseline showing the position of S148. A complete analysis of S148 would exceed the scope of this work. However, we find that S148 is following the disk structure of the S-cluster as proposed by \cite{Ali2020}.}
\label{tab:orbit_elements_s148}
\end{table*}
As mentioned in the previous section (Sec. \ref{sec:s4716_2016_2017_2018}), the star shows a variable magnitude between 2016 and 2018. For the data of 2016, we measure a K-band magnitude of 16.9$\pm$0.1 mag for S148. The star gets significantly brighter by almost 1 mag in 2017, resulting in a K-band magnitude of 16.0$\pm$0.1 mag.
This is followed by a decreased K-band magnitude to 17.0$\pm$0.1 in 2018 for S148. We can safely conclude that S4716 combined with S148 creates a so-called blend star \citep[][]{Sabha2012, Eckart2013}. This apparent star is the result of at least 2 fainter stars and shows a higher K-band magnitude compared to the individual S-stars S148 and S4716. The created blend star is an example for the checklist introduced in the next section and underlines the need for longer data baselines. A detailed analysis of this event and a comprehensive analysis of this star exceeds the scope of this work. However, we will study S148 and the corresponding blend star event in detail in an upcoming publication (Pei\ss{}ker et al., in prep.).


\section{Mass relation in the S-cluster}
\label{sec:mass_estimator}
\begin{figure}[htbp!]
	\centering
	\includegraphics[width=.5\textwidth]{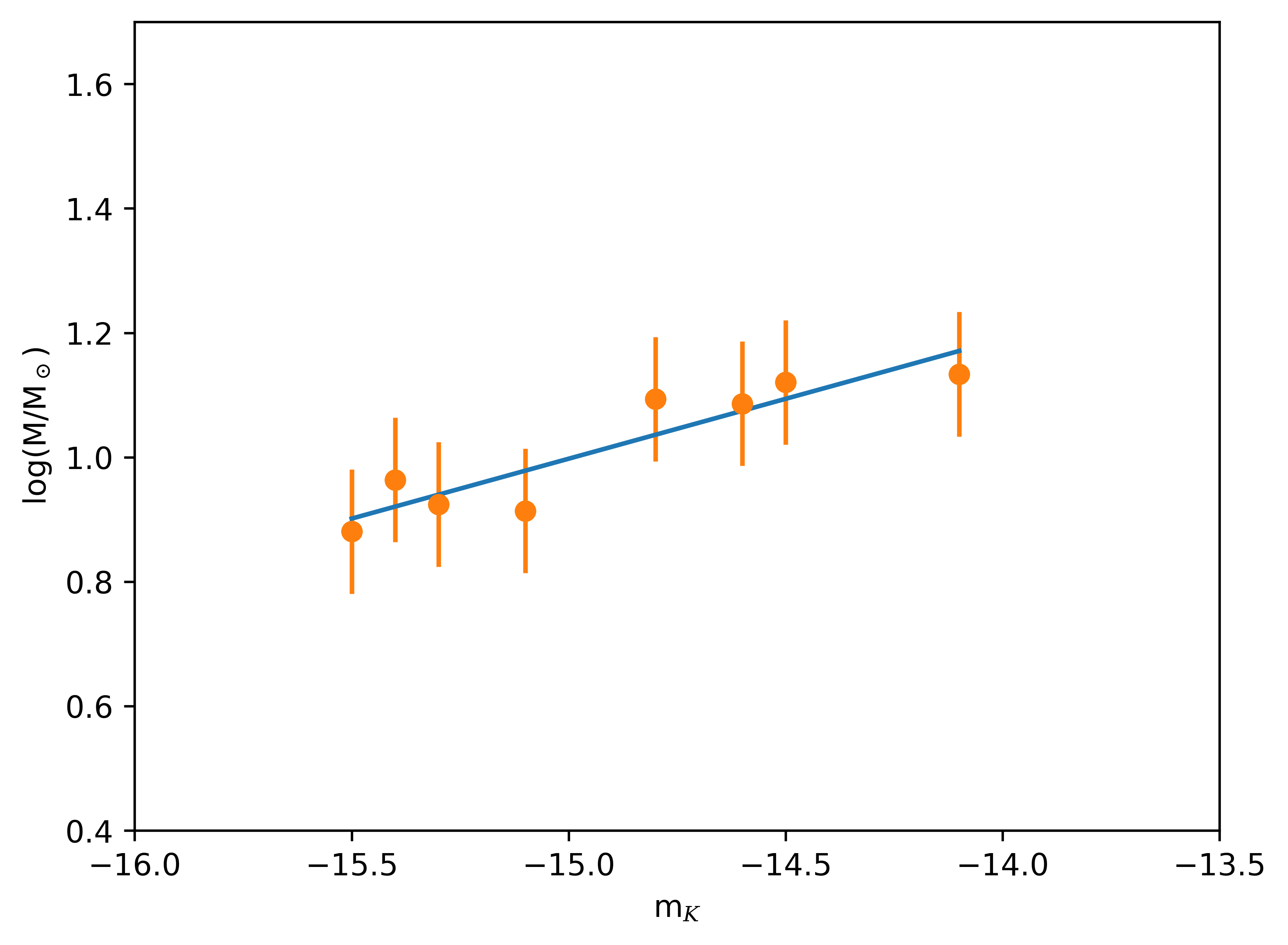}
	\caption{Relation between $\rm log(M/M_{\odot})$ and the K-band magnitude m$_K$. We use the estimated mass and magnitude of \cite{Habibi2017}.}
\label{fig:habibi_data}
\end{figure}

Here we outline an accessible method for deriving the mass of the stellar members of the S-cluster. For this, we use the estimated mass and K-band magnitudes of \cite{Habibi2017}. The authors of Habibi et al. use a 12 year data baseline achieved with SINFONI. We follow the inspired analysis of \cite{Peissker2020d} and fit a 1 dimensional polynom to the derived mass and magnitude of the S-stars (Fig. \ref{fig:habibi_data}).
We use $log\frac{M}{M_{\odot}}\,=\,m\cdot x+b$ where x denotes the K-band magnitude m$_K$. With the data from \cite{Habibi2017} and Fig. \ref{fig:habibi_data}, we find $log\frac{M}{M_{\odot}}\,=\,0.1925\cdot m_K +3.885$.

\section{Checklist for the data}
\label{appendix:checklist}

Here we will list some key aspects to verify findings with the here presented approach using high-pass filter for the analysis. Of course, this list can be applied to previous and future publications.
\begin{enumerate}
    \item Independent observations: if possible, one should use several instruments/telescopes.
    \item Long term: to avoid blend star scenarios, at least 3 consecutive years of observations should exclude this possibility because of the high proper motion of the S-cluster members.
    \item Spectrum: the Kepler fit arranges a possible orbit in the 3 dimensional space based on data points observed on the projected sky. Hence, the result of the fit also delivers a solution for the LOS velocity. If possible, a spectrum of the observed star can be used to verify the finding as an independent parameter.
    \item Inside a reasonable uncertainty range, the proper motion should coincide with the observation of the star.
    \item The latter bullet point can also be applied to the magnitude.
    \item Because of the stellar density of the cluster, unidentified stars can be located at the airy rings of known S-stars. If the star can only be detected on the airy rings, it is reasonable to assume that the finding is an artefact.
\end{enumerate}


\end{document}